\documentclass[aps,twocolumn,showpacs,amsmath,amssymb,nofootinbib,superscriptaddress,showkeys]{revtex4-1}

\usepackage{epsfig}
\usepackage{slashed}

\makeatletter
\def\slashchar#1{{\mathpalette\c@ncel{#1}}} % TeXbook, bottom of p360
\makeatother
\def\vsl{\slashchar{v}}
\begin{document}

\title{The Heavy Quark Spin Symmetry Partners of the $X(3872)$}
 \author{J. Nieves}

 \author{M. Pav\'on Valderrama}\email{m.pavon.valderrama@ific.uv.es}
 \affiliation{Instituto de F\'{\i}sica Corpuscular (IFIC), Centro
   Mixto CSIC-Universidad de Valencia, Institutos de Investigaci\'on
   de Paterna, Aptd. 22085, E-46071 Valencia, Spain}

\date{\today}

\begin{abstract}
\rule{0ex}{3ex}
We explore the consequences of heavy quark spin symmetry
for the charmed meson-antimeson system in a contact-range
(or pionless) effective field theory.
As a trivial consequence, we theorize the existence of a heavy quark
spin symmetry partner of the $X(3872)$, with $J^{PC}=2^{++}$, which we
call $X(4012)$ in reference to its predicted mass.
If we additionally assume that the $X(3915)$ is a $0^{++}$ heavy spin symmetry
partner of the $X(3872)$, we end up predicting a total of six
$ D^{(*)}\bar{D}^{(*)}$ molecular states.
We also discuss the error induced by higher order effects such as 
finite heavy quark mass corrections, pion exchanges and
coupled channels, allowing us to estimate the expected
theoretical uncertainties in the position of these new states.
\end{abstract}

% Optional PACS: 03.65.Ge (bound states), 03.65.Nk (scattering)
\pacs{03.65.Ge, 13.75.Lb, 14.40.Lb, 14.40.Pq, 14.40.Rt}

\maketitle

\section{Introduction}

The discovery of the $X(3872)$ resonance~\cite{Choi:2003ue} might have
confirmed a well-known theoretical expectation of hadronic physics,
heavy meson molecules~\cite{Voloshin:1976ap,Tornqvist:1991ks,
Tornqvist:1993ng,Manohar:1992nd,Ericson:1993wy}.
The $X(3872)$, with a mass of $m_X = 3871.57 \pm 0.25\,{\rm
MeV}$~\cite{Nakamura:2010zzi}, is extremely close to the
$D^{0*}\bar{D}^0$ threshold ($m_{D^0} + m_{D^{*0}} = 3871.73 \pm
0.21 \,{\rm MeV}$), a feature suggesting that its nature is mostly
molecular, where the role of other more compact components
(e.g. tetraquark, $c\bar{c}$) will be less important in comparison.
The $J^{PC}$ quantum numbers of the $X(3872)$ are either $1^{++}$ or
$2^{-+}$~\cite{Abe:2005iya,Abulencia:2006ma,delAmoSanchez:2010jr},
with a slight experimental preference for the second set of quantum
numbers~\cite{delAmoSanchez:2010jr}. However, only the $1^{++}$
assignment is compatible with the molecular interpretation.
If the $2^{-+}$ possibility turns out to be the correct one,
the $X(3872)$ would be much more exotic than expected.
In this regard, the analysis of Hanhart et al.~\cite{Hanhart:2011tn}
suggests that the experimental data, though too scarce to draw
definitive conclusions, might indeed be more compatible with $1^{++}$ after all
(see, however, Ref.~\cite{Faccini:2012zv} for a different opinion).
Meanwhile, in the bottom sector,
the recent discovery of the $Z_b(10610)$ and $Z_b(10650)$ isovector
states by Belle~\cite{Collaboration:2011gj,Belle:2011aa}
also provides two new strong candidates for molecular states,
as the $Z_b$'s lie close to the $B^*\bar{B}$ and $B^*\bar{B}^*$
thresholds respectively. 

Actually, not all the heavy meson molecules are necessarily
as shallow as the $X(3872)$ or the two $Z_b$ resonances.
Several theoretical works have suggested the molecular interpretation
of other XYZ states: the $X(3915)$~\cite{Uehara:2009tx} and
the $Y(4140)$~\cite{Aaltonen:2009tz} have been identified
as a $ D^*\bar{D}^*$ and $ D_s^*\bar{D}_s^*$ bound state respectively
by different theoretical approaches~\cite{Liu:2009ei,Branz:2009yt,Ding:2009vd}.
The $Y(4260)$~\cite{Aubert:2005rm} has even been proposed to have
the three body structure $J/\Psi K \bar{K}$~\cite{MartinezTorres:2009xb}.
Moreover other states have been predicted,
especially in the isoscalar bottom
sector~\cite{Liu:2008tn,Sun:2011uh,Nieves:2011zz},
but have not yet been experimentally confirmed or discarded.

Heavy meson molecules are a natural thing to expect % if we consider
on the basis of the similarity between the meson-meson interaction and
the nuclear force that binds
the deuteron~\cite{Voloshin:1976ap,Tornqvist:1993ng}.
From this analogy, we expect that the effective field theory (EFT)
formulation of nuclear physics~\cite{Beane:2000fx,Bedaque:2002mn,Epelbaum:2005pn,Epelbaum:2008ga,Machleidt:2011zz}
will also represent a constructive approach to the description of
heavy meson systems at low-energies.
As in the nucleon-nucleon system, the low energy interaction between
a pair of heavy mesons is mediated by pion exchanges, which in turn
are constrained by chiral symmetry.
In contrast, the nature of the short range interaction remains unknown,
but we can parametrize it in terms of contact-range operators
between the nucleon / heavy meson fields. 
However, in the case of heavy meson molecules
there is a particularly simplifying feature:
pion exchanges are weaker than in the nuclear case, owing to the smaller
light quark content of the heavy mesons in comparison to the nucleons.
This means that pions are amenable to a perturbative treatment
in a larger range of energies than in the nucleon-nucleon system.
From this it is expected that the EFT description of heavy meson
molecules will simplify at lowest order to a contact range theory,
at least for a certain binding energy window~\cite{Valderrama:2012}.
A nice illustration of this idea is provided by X-EFT~\cite{Fleming:2007rp},
which considers the low energy description of the $X(3872)$ state
as a $D^0\bar{D}^{0*}$ molecule.

On a different level, the presence of a charm quark/antiquark
in the heavy meson and antimeson conforming the $X(3872)$ dictates
that heavy quark spin symmetry (HQSS)~\cite{Isgur:1989vq,Isgur:1989ed,Neubert:1993mb,Manohar:2000dt}
is relevant for this system.
In the context of the EFT description of heavy meson molecules,
HQSS constrains the form of the contact range operators
of the theory in a very specific way~\cite{AlFiky:2005jd}.
What this means is that there should be HQSS partners of the $X(3872)$,
in analogy with the theorized HQSS partners of the $Z_b(10610)$
and $Z_b(10650)$ that have been already discussed
in Refs.~\cite{Voloshin:2011qa,Mehen:2011yh}.
The purpose of this paper is to investigate the HQSS structure of charm
meson-antimeson (${\rm D}$, ${\rm D}^{*}$ and ${\rm \bar D}$,
${\rm \bar D}^{*}$) molecules, with the intention of
identifying the possible HQSS partners
of the $X(3872)$.
For that we will assume that certain XYZ states are molecular,
in particular the $X(3915)$~\cite{Uehara:2009tx}.
This identification will span a total of six states,
some of which have also been predicted
in other schemes.

The article is structured as follows: in Sect.~\ref{sec:EFT-LO}
we present the EFT description of heavy meson molecules that 
we advocate, which is in turn inspired
on the ideas of Ref.~\cite{Valderrama:2012}.
We explore the consequences of this EFT in Sect.~\ref{sec:partners},
where we deduce from HQSS the existence of the $2^{++}$
$ D^*\bar{D}^*$ partner of the $X(3872)$,
and of other four additional states if we identify the $X(3915)$ as a
$0^{++}$ $ D^*\bar{D}^*$ molecule.
In Sect.~\ref{sec:subleading} we probe the robustness of the previous
results by considering the effect of subleading order contributions,
namely finite heavy quark mass corrections, pion exchanges
and particle coupled channel effects. In Sect.~\ref{sec:decay}, we
briefly discuss some HQSS constraints on the total decay widths of the
states found in this work.
In Sect.~\ref{sec:conclusion} we discuss the significance of the results.
Finally, the technical details of the manuscript can be consulted in Appendices
\ref{app:EFT}, \ref{app:projecting} and \ref{app:LO-pot}.

\section{The EFT Description at Lowest Order}
\label{sec:EFT-LO}

In this section we review the EFT framework that we use for the description
of heavy meson-antimeson molecules at lowest order.
The presentation is simple and schematic, centered in the conceptual issues
rather than the specific details, which can be consulted
in Ref.~\cite{Valderrama:2012}.
The EFT we are advocating is in fact identical to the one presented
in Ref.~\cite{Mehen:2011yh} for the isovector bottom sector.
Following the findings of Ref.~\cite{Valderrama:2012},
we assume pion exchanges and particle coupled channel effects
to be a subleading correction entering at next-to-leading order
(${\rm NLO}$) and next-to-next-to-leading order (${\rm N^2LO}$)
respectively.
Nevertheless, we will perform explicit calculations
to test these assumptions.

\subsection{Overview of the EFT Formalism}

The EFT approach provides the possibility of constructing generic and
systematic descriptions of arbitrary low energy processes.
They are particularly useful when the system we are interested in
cannot be easily explained in terms of a more fundamental
description at higher energies.
The EFT idea is simple: we identify the fields and symmetries that
are relevant at low energies and construct all possible
interactions compatible with them.
Even though the number of interactions is infinite,
they can be classified according to their importance at low energies
by means of power counting, the ordering principle of EFT.
If $Q$ is the soft (low energy) scale of the system we are describing
and $\Lambda_0$ the hard (high energy) scale, power counting allows
us to express any physical quantity as a power series
in terms of the small parameter $x_0 = Q / \Lambda_0$.

For illustrating this idea let us consider a physical quantity $A$
that we want to compute in the EFT framework.
This quantity receives in principle contributions from all the relevant
diagrams involving the low energy fields and compatible with the low
energy symmetries:
\begin{eqnarray}
A (Q, \Lambda_0) = \sum_{D} A^{(D)} (Q, \Lambda_0) \, .
\end{eqnarray}
However, the different diagrams have different scaling properties that
can be used for ordering the sum above.
For example, we have the canonical dimension of $A$, which is defined as
\begin{eqnarray}
\label{eq:scaling_dim}
A^{(D)} (\lambda\,Q, \lambda\,\Lambda_0) = \lambda^{d_A} \,
A^{(D)}(Q, \Lambda_0) \, ,
\end{eqnarray}
and is the same for all the EFT contributions to $A$.
But the interesting scaling property is power counting, which refers to
the behaviour under a transformation of the type $Q \to \lambda \, Q$
\begin{eqnarray}
\label{eq:scaling_count}
A^{(D)} (\lambda\,Q, \Lambda_0) = \lambda^{\nu_D} \, A^{(D)}(Q, \Lambda_0) \, ,
\end{eqnarray}
where $\nu_D$ is the order of the contribution $D$, which is
bounded from below (i.e. $\nu_D \geq \nu_0$).
The sum of diagrams above can be reorganized as an expansion
in terms of increasing scaling dimension:
\begin{eqnarray}
A (Q, \Lambda_0) = \sum_{\nu \geq \nu_0} A^{(\nu)} (Q, \Lambda_0) \, ,
\end{eqnarray}
where, for simplicity, we have get rid of the $D$ subscripts and
superscripts.
For each order $\nu$ there is only a finite number of diagrams
that contributes to the quantity $A$.
Combining the scaling laws of Eqs.~(\ref{eq:scaling_dim})
and (\ref{eq:scaling_count}) we obtain a well-defined
power series for $A$
\begin{eqnarray}
A(Q, \Lambda_0) &=& \Lambda_0^{d_A} \,
\sum_{\nu \geq {\rm \nu_0}} \,
{\left( \frac{Q}{\Lambda_0} \right)}^{\nu}\,\hat{A}^{(\nu)}(\frac{Q'}{Q}) \nonumber \\
&=& \Lambda_0^{d_A} \,\sum_{\nu \geq {\rm \nu_0}}\,
\, x_0^{\nu}\,\hat{A}^{(\nu)}(\frac{Q'}{Q}) \, ,
\end{eqnarray}
with $\hat{A}^{(\nu)}$ a dimensionless quantity that we expect to
be of the order of unity (i.e. $Q^0$).
Notice that $\hat{A}^{(\nu)}$ does not depend on the hard scale $\Lambda_0$ 
and is related to $A^{(\nu)}$ via Eqs.~(\ref{eq:scaling_dim})
and (\ref{eq:scaling_count}).
In the formula above, $Q'$ is an auxiliary soft scale we use to express
$\hat{A}^{(\nu)}$ as a function of a dimensionless ratio.
Provided there is a clear scale separation in the system, that is,
$\Lambda_0 \geq Q$, the power series above will be convergent.
Not only that, if we consider only contributions from diagrams
with $\nu < \nu_{\rm max}$, the error of the EFT calculation
will be $x_0^{\nu_{\rm max} + 1}$.
In this work we will only perform calculations at the lowest order
and we expect a relative error of the order of $x_0$ in the
calculations to follow.

If we are interested in the low energy description of heavy meson-antimeson
bound states, the relevant physical object we want to expand is the
(non-relativistic) potential between the heavy meson and antimeson:
\begin{eqnarray}
\label{eq:V-expansion}
V &=& \sum_{\nu = \nu_0}^{\nu_{\rm max}} V^{(\nu)} +
\mathcal{O}(x_0^{\nu_{\rm max} + 1}) \, .
\end{eqnarray}
The expansion starts at order $\nu_0 \geq -1$, where $x_0$ is the ratio of
the soft and hard scales of the system.
The low energy degrees of freedom we consider are the heavy meson and
antimeson fields and the pion field.
The pion-meson vertices are constrained by chiral symmetry
and the corresponding Feynman rules can be derived from
heavy hadron chiral perturbation theory (HHChPT)~\cite{Wise:1992hn}.
In turn HQSS generates strong constraints on the form of the heavy
meson-antimeson interactions~\cite{AlFiky:2005jd}.
This means that the EFT potential includes two kind of contributions:
contact range interactions, i.e. four meson vertices,
and pion exchanges.
The set of soft scales $Q$ includes in principle the pion mass $m_{\pi}$
and the center-of-mass momenta $\vec{p}$ and $\vec{p}\,'$ of
the meson and antimeson.
The hard scale $\Lambda_0$ can represent the momentum scale at which
we expect the low energy symmetries to break down, i.e.
the chiral symmetry breaking scale
$\Lambda_{\chi} = 4 \pi f_{\pi} \sim 1\,{\rm GeV}$
(with $f_{\pi}$ the pion decay constant) for chiral symmetry
and the heavy quark mass $m_Q$ for HQSS,
but it can also stand for the momentum scale at which the composite structure
of the heavy mesons starts to be resolved.

At lowest (or leading) order ($Q^0$) the heavy meson-antimeson EFT potential
is local and only depends on the momentum exchanged by
the meson and anti-meson
\begin{eqnarray}
\langle \vec{p}\,| V^{(0)} | \vec{p}\,'\, \rangle 
= V^{(0)}(\vec{p} - \vec{p}\,'\,) \, ,
\end{eqnarray}
where we have
\begin{eqnarray}
\label{eq:V-EFT-lowest-order}
V^{(0)}(\vec{q}\,) &=& C_0^{(0)} + \eta \frac{g^2}{2
  f_{\pi}^2}\, \frac{(\vec{a} \cdot \vec{q}\,)\,(\vec{b} \cdot
  \vec{q}\,)}{\vec{q}^{\,\,2} + m_{\pi}^2} \, . \label{eq:OPEtype}
\end{eqnarray}
As can be seen, the potential is the sum of a contact
and a finite range contribution.
The contact range operator $C_0$ is a free parameter of the theory.
The finite range contribution is the well-known one pion exchange (OPE)
potential, where $g \simeq 0.6$ is the axial coupling between
the heavy meson and the pion (we have particularized its value
for the charmed meson case,
see Refs.~\cite{Ahmed:2001xc,Anastassov:2001cw} for a determination),
$f_{\pi} \simeq 132\,{\rm MeV}$ the pion decay constant,
and $\vec{q}$ the momentum exchanged by the heavy meson
and antimeson.
The sign $\eta$ and the polarization operators $\vec{a}$ and $\vec{b}$ 
depend on whether the initial/final states is ${\rm P\bar{P}}$,
${\rm P\bar{P}^*}$, ${\rm P^*\bar{P}}$ or ${\rm P^*\bar{P}^*}$. 
A more detailed account can be found in Appendices \ref{app:EFT},
\ref{app:projecting} and \ref{app:LO-pot},
where the leading order (${\rm LO}$) potential is derived.

A problem with the EFT potential above is its behaviour for larges values of
the exchanged momentum, $|\vec{q}\,| \gg m_{\pi}$.
In this limit the ${\rm LO}$ potential tends to a constant value.
At higher orders in the EFT expansion the problem worsens and
the potential actually diverges.
This feature can be easily deduced from power counting
\begin{eqnarray}
\label{eq:V-EFT-counting}
V^{(\nu)}(\lambda Q) = \lambda^{\nu} V^{(\nu)}(Q) \, ,
\end{eqnarray}
which admits a solution of the type
$V^{(\nu)}(\vec{q}\,) \propto |\vec{q}\,|^{\nu}$ for $|\vec{q}\,| \gg m_{\pi}$.
Of course, if we are only considering tree level amplitudes involving
no heavy meson-antimeson loops, then there is no conceptual problem
with the previous divergences: the EFT potential is only expected
to make sense at low energies.
However, if we iterate the potential in the Sch\"odinger or Lippmann-Schwinger
equation (a necessary step for the description of bound states)
the divergences will require renormalization.

In non-relativistic EFTs the renormalization process is straightforward.
First, we begin by regularizing the potential:
\begin{eqnarray}
\label{eq:V-EFT-regularization}
\langle \vec{p}\, | V_{\Lambda} | \vec{p}\,' \rangle = 
f(\frac{\vec{p}}{\Lambda})\,
\langle \vec{p}\, | V | \vec{p}\,' \rangle \,
f(\frac{\vec{p}\,'}{\Lambda}) \, ,
\end{eqnarray}
where $V_{\Lambda}$ is the regularized potential that we will employ
in actual calculations, $V$ is the unregularized (i.e. the original)
potential, $f(x)$ a regulator function and $\Lambda$
an ultraviolet cut-off.
In the calculations to follow, we will use a gaussian regulator
of the type $f(x) = e^{-x^2}$.
At this point physical predictions can still depend strongly on the value
of the cut-off, a problem we must solve.
Thus there is a second step in the renormalization process:
we allow the contact range operators to depend on the cut-off.
At lowest order this implies that $C^{(0)} = C^{(0)}(\Lambda)$.
If we have included all the counterterms required by power counting,
they will be able to absorb all the divergences of the theory.
The calculations will still contain a residual cut-off dependence,
but this does not represent a problem: its size is expected
to be a higher order effect, at least for a judicious choice
of the cut-off.
Even though there is not a well-established criterion for choosing the cut-off,
calculations in nuclear EFT suggest that the optimal value of the cut-off
should never be much larger than the hard scale
$\Lambda_0$~\cite{Epelbaum:2009sd}.
Here, for avoiding the breakdown of the low energy symmetries in loops,
we advocate the slightly more stringent condition $\Lambda \leq \Lambda_0$.
We employ the actual values $\Lambda = 0.5\,{\rm GeV}$ and $1\,{\rm GeV}$,
for which we have checked that the results do not change considerably.
This indicates that renormalization has been correctly implemented.

\subsection{Power Counting and Bound States}

The description of bound states requires the iteration of
the EFT potential in the bound state equation
\begin{eqnarray}
\label{eq:bs-eq}
| \Psi_B \rangle = G_0(E) V | \Psi_B \rangle \, ,
\end{eqnarray}
where $| \Psi_B \rangle$ is the wave function, $G_0(E) = 1 / (E - H_0)$
the resolvent operator and $V$ the potential.
If we require the bound state equation to be compatible with the power
counting of the EFT potential, then successive iterations of
the $G_0 V$ combination must be of the same order:
\begin{eqnarray}
\mathcal{O}\,(V) = \mathcal{O}\,(V G_0 V) \, .
\end{eqnarray}
By taking into account that the $G_0$ operator scales as $Q$
in loops~\footnote{This can be trivially checked by considering
the rescaling transformation
$$
\int \frac{d^3\vec{q}}{(2 \pi)^3}\,G_0(\lambda^2 E)
= \lambda \int \frac{d^3 \vec{q}}{(2 \pi)^3}\,G_0(E) \, ,
$$
where the energy rescales as $\lambda^2$ as we are considering
a non-relativistic system.},
we can see that only the order $Q^{-1}$ contribution to the potential
should be iterated.
Thus the presence of shallow bound states or large scattering lengths
in a two-body system requires the non-perturbative treatment of
a piece of the effective potential.
There exists a problem then: the EFT potential we obtain from HHChPT starts
at order $Q^0$ and is therefore incompatible with the EFT description
of a low energy bound state.

The solution is to redefine power counting by promoting
the $C_0$ contact range operator from order
$Q^0$ to $Q^{-1}$~\cite{Kaplan:1998tg,Kaplan:1998we,Gegelia:1998gn,Birse:1998dk,vanKolck:1998bw}.
This is equivalent to assume that the $C_0$ operator is contaminated
by a low energy scale. We can confirm this assumption {\it a
  posteriori} by solving the bound state equation with the $C_0$
operator alone:
if we regularize the EFT potential with a cut-off $\Lambda$, and set
the value of $C_0(\Lambda)$ to reproduce the position of
the bound state, we obtain the generic result 
\begin{eqnarray}
\label{eq:C0-running}
\frac{1}{C_0(\Lambda)} \sim \frac{\mu}{2\pi}\,
\left( \gamma_B - \frac{2}{\pi}\,\Lambda\right) \, ,
\end{eqnarray}
where $\mu$ is the reduced mass of the two-body system and
$\gamma_B = \sqrt{-2 \mu E_B}$ the wave number
(with $E_B$ the bound state energy).
Of course, the exact form of the relation depends on the specific
regularization scheme employed -- the expression above corresponds to
power divergence subtraction~\cite{Kaplan:1998tg,Kaplan:1998we} --
but will be in the line of the previous form.
From the equation above it is evident that we can only have 
$C_0 \sim Q^{-1}$ as long as the cut-off $\Lambda$ scales
like ${\cal O}(Q)$. 
Note that this is consistent with our requirement that all involved momenta
should be soft and smaller than the charm quark mass to make sense of HQSS.

\subsection{The EFT Potential at Lowest Order}

As we have seen,
the $S$-wave ${\rm LO}$ interaction (${\cal O}(Q^{-1})$) between a heavy meson
(${\rm H = P, P^*}$) and antimeson (${\rm {\bar H} = {\bar P}, {\bar P}^*}$)
only contains contact operators (i.e. four heavy meson vertices).
The contact range interactions are in turn constrained by HQSS and
depend on the particle channel under consideration
(${\rm H\bar{H}} = {\rm D\bar{D}}, {\rm D\bar{D}^*}/{\rm D^*\bar{D}},
{\rm D\bar{D}}$),
the total isospin $I$ and the value of the $J^{PC}$ quantum numbers.
In particular HQSS limits the number of independent ${\rm LO}$
counter-terms to two per isospin channel at ${\rm LO}$~\cite{AlFiky:2005jd}.
We will only consider the isoscalar $I = 0$ channels, to which the
$X(3872)$ and the $X(3915)$ belong,
in the isospin symmetric limit.

The HQSS contact range interaction can mix different particle channels
with the same $J^{PC}$ quantum numbers.
Therefore, for writing the ${\rm LO}$ potential we considering
the set of particle coupled channel basis
\begin{eqnarray}
\mathcal{B}(0^{++}) &=&
\left\{ | P\bar{P} \rangle , | P^*\bar{P}^* (0) \rangle \right\} \, , \\
\mathcal{B}(1^{+-}) &=&
\left\{ \frac{1}{\sqrt{2}}\left( | P\bar{P}^* \rangle + | P^*\bar{P} \rangle
\right) , | P^*\bar{P}^*(1) \rangle \right\} \, , \\
\mathcal{B}(1^{++}) &=&
\left\{ \frac{1}{\sqrt{2}}\left( | P\bar{P}^* \rangle - | P^*\bar{P} \rangle
\right) \right\} \, , \\
\mathcal{B}(2^{++}) &=&
\left\{ | P^*\bar{P}^*(2) \rangle \right\} \, ,
\end{eqnarray}
where the number in parenthesis in the $|P^*\bar{P}^*(S) \rangle$ states 
is the total intrinsic spin $S$ to which the vector meson-antimeson
system couples.
In this basis, the EFT potential is independent of momentum and
reads (see Appendices \ref{app:EFT},\ref{app:projecting} and \ref{app:LO-pot})
\begin{eqnarray}
V^{\rm LO}(\vec{q}, {0^{++}}) &=& 
\begin{pmatrix}
C_{0a} & \sqrt{3}\,C_{0b} \label{eq:contact1}\\
\sqrt{3}\,C_{0b} & C_{0a} - 2\,C_{0b}
\end{pmatrix} \, , \\
V^{\rm LO} (\vec{q}, 1^{+-}) &=&
\begin{pmatrix}
C_{0a} - C_{0b} & 2\,C_{0b} \\
2\,C_{0b} & C_{0a} - C_{0b}
\end{pmatrix} \, , \\
V^{\rm LO}(\vec{q}, 1^{++}) &=& C_{0a} + C_{0b} \label{eq:contact2-a} \, , \\
V^{\rm LO}(\vec{q}, 2^{++}) &=& C_{0a} + C_{0b} \label{eq:contact2-b} \, ,
\end{eqnarray} 
where $C_{0a}$ and $C_{0b}$ are the two independent counter-terms
that we expect from HQSS for a given isospin sector.
This contact potential $V$ behaves as ${\cal O}(Q^{-1})$, but for avoiding
confusions with the notation we have renamed $V^{(-1)}$ to $V^{\rm LO}$,
as the negative exponent could be confused by
the inverse of the potential. 
As can be appreciated, the $1^{++}$ and $2^{++}$ cases are uncoupled and
their potential is identical, a strong hint that we should expect
a $2^{++}$ $\rm D^*\bar{D}^*$ partner of the $X(3872)$.

At this point we notice that the heavy pseudoscalar and vector mesons
$\rm P$ and $\rm P^{*}$ are only degenerate in the heavy quark limit
$m_Q \to \infty$.
For finite $m_Q$ there is a mass splitting between the heavy mesons
\begin{eqnarray}
M_{P^*} - M_{P} = \Delta_Q \, ,
\end{eqnarray}
that scales as $1/m_Q$.
As a consequence of this gap, the two $\rm  H\bar{H}$ thresholds 
in the $0^{++}$ and $1^{+-}$ coupled channel happen at different energies.
If we are interested in low-lying bound states, the energy difference
between the two thresholds may actually be considerably larger than
the binding energy of the state.
Within the EFT framework this means that we may very well be entitled
to ignore the coupled channel effects.
The momentum scale associated with the coupled channels is
\begin{eqnarray}
\Lambda_{C}(0^{++}) &=& \sqrt{2\mu\,(2\Delta_Q)} \, , \\
\Lambda_{C}(1^{+-}) &=& \sqrt{2\mu\,\Delta_Q} \, ,
\end{eqnarray}
for the $0^{++}$ and $1^{+-}$ cases respectively, where $\mu$ is
the reduced mass of the $\rm  H\bar{H}$ heavy meson system and
$\Delta_Q$ is the energy gap.
In the particular case of the charm mesons, direct evaluation yields
$\Lambda_{C(0^{++})} \sim 750\,{\rm MeV}$ and $\Lambda_{C(1^{+-})}
\sim 520\,{\rm MeV}$.
If we have a $0^{++}$ $\rm D\bar{D}$ ($1^{+-}$ $\rm D^*\bar{D}$)
bound state at threshold, the corresponding $0^{++}$ $\rm D^*\bar{D}^*$
($1^{+-}$ $\rm D^*\bar{D}^*$) component will have a wave number at least
of the order of the hard scale $\Lambda_0 \sim 0.5-1\,{\rm GeV}$.
Thus there is no problem in ignoring the coupled channel structure and
treating the two particle channels in the $0^{++}$ and $1^{+-}$ cases
as independent.
From a more formal EFT viewpoint what we are doing is to consider
the coupled channel momentum scale as $\Lambda_C \sim {\cal O}(Q^0)$,
from which  we expect the $G_0$ operator involved in the particle
mixing  to scale like $Q^3$ (see for instance Ref.~\cite{Valderrama:2012}).
This translates into a suppression of particle coupled channel effects
by two orders in the chiral expansion:
if we count the $C_{0b}$ operator as $Q^{-1}$, then particle coupled channels
do not enter until order $Q$, that is, at least one order
beyond pion exchanges.

Taking into account that we can distinguish between different particle
channels in the EFT, the ${\rm LO}$ potentials for the $0^{++}$
and $1^{+-}$ cases finally simplify to
\begin{eqnarray}
V^{\rm LO}_{{\rm P \bar P}}(\vec{q}, {0^{++}}) &=& C_{0a} \label{eq:coni} \, , \\
V^{\rm LO}_{{\rm P^* {\bar P}^* }}(\vec{q}, {0^{++}})
&=& C_{0a} - 2\,C_{0b} \, , \\
V^{\rm LO}_{{\rm P^* {\bar P} / P {\bar P}^*}}(\vec{q}, {1^{+-}})
&=& C_{0a} - C_{0b} \, , \\ 
V^{\rm LO}_{{\rm P^* {\bar P}^* }}(\vec{q}, {1^{+-}})
&=& C_{0a} - C_{0b} \, , \label{eq:conf} 
\end{eqnarray}
where we have added the particle channel as a subscript for distinguishing
states with the same $J^{PC}$ quantum numbers.
Even though the particle coupled channel structure of the $0^{++}$ and $1^{+-}$
molecules can be ignored in the ${\rm LO}$ description of these states,
we expect the $0^{++}$ and $1^{+-}$ $\rm P^*\bar{P}^*$ states
to have a strong tendency to decay to $\rm P\bar{P}$ and
$\rm P\bar{P}^*$ respectively.
On the contrary, the $2^{++}$ $\rm P^*\bar{P}^*$ will have a smaller partial
decay width to a heavy meson-antimeson pair, as this process does not happen
via a $Q^{-1}$ counterterm, and therefore will be suppressed by one order
in the EFT expansion.

\subsection{Bound States at Lowest Order}

Finally, for completeness, we briefly discuss the solution of
the bound state equation for the contact range potentials
that appear in the ${\rm LO}$ description of
heavy molecular states.
Even though well-known, it will be of help
for the calculations in the next section.
We solve the bound state equation with the regularized potential $V_{\Lambda}$
that reads
\begin{eqnarray}
\label{eq:V-LO-regularization}
\langle \vec{p} \, | V^{\rm LO}_{\Lambda} | \vec{p}\,'\, \rangle =
f(\frac{\vec{p}}{\Lambda})\,C_0(\Lambda)\,f(\frac{\vec{p}}{\Lambda})
\, ,
\end{eqnarray}
where $f(x)$ is the regulator, and $C_0$ the appropriate counterterm
for each $J^{PC}$ / particle channel combination.
For this potential, the wave function admits the ansatz
\begin{eqnarray}
\langle \vec{p} \, | \Psi_B \rangle = \mathcal{N}\,
\frac{2\mu}{p^2 + \gamma^2}\,f(\frac{\vec{p}}{\Lambda}) \, ,
\end{eqnarray}
where $\mathcal{N}$ is a normalization constant, $\mu$ the reduced mass
of the two body system and $\gamma = \sqrt{-2\mu E_B}$ the wave number
of the state, with $E_B$ the binding energy.
Direct substitution into the bound state equation (Eq.~(\ref{eq:bs-eq}))
yields
\begin{eqnarray}
-\frac{1}{C_0(\Lambda)} = \int \frac{d^3 \vec{q}}{(2 \pi)^3}\,
f^2(\frac{\vec{q}}{\Lambda}) \, \frac{2 \mu}{q^2 + \gamma^2} \, .
\end{eqnarray}
This is the eigenvalue equation of the contact range theory.
For the $\rm H\bar{H}$ system there are six of these equations
corresponding to the six possible S-wave states: in each case,
we simply need to particularize the values of $C_0$ and $\mu$
as appropriate.
For further details we refer the reader to Ref.~\cite{Nieves:2011zz},
where the equations above were derived and discussed
in the context of heavy meson-antimeson molecules.

\section{The Partners of the $X(3872)$}
\label{sec:partners}

\begin{table*}
\begin{center}
\begin{tabular}{|c|c|c|c|c|c|c|}
\hline \hline
$J^{PC}$ & $\rm H\bar{H}$ & $^{2S+1}L_J$  & $V_C$
& $E$ $(\Lambda =0.5$ GeV) & $E$ $(\Lambda =$ 1 GeV) & ${\rm
  Exp}$~\cite{Nakamura:2010zzi} \\
\hline
$0^{++}$ & $ D\bar{D}$ & $^1S_0$ & $C_{0a}$
& $3706 \pm 10$ & $3712^{+13}_{-17} $  & $-$ \\
\hline
$1^{++}$ & $ D^*\bar{D}$ & $^3S_1$ & $C_{0a} + C_{0b}$
& Input & Input & $3872$ \\
$1^{+-}$ & $ D^*\bar{D}$ & $^3S_1$ & $C_{0a} - C_{0b}$
& $3814 \pm 17$ & $3819^{+24}_{-27}$ & $-$ \\
\hline
$0^{++}$ & $ D^*\bar{D}^*$ & $^1S_0$ & $C_{0a} - 2\,C_{0b}$ & Input
& Input & $3917$ \\
$1^{+-}$ & $ D^*\bar{D}^*$ & $^3S_1$ & $C_{0a} - C_{0b}$
& $3953\pm 17$ & $3956^{+25}_{-28}$ & $3942$ \\
$2^{++}$ & $ D^*\bar{D}^*$ & $^5S_2$ & $C_{0a} + C_{0b}$
& $4012 \pm 3$ & $4012^{+4}_{-9} $ & $-$ \\
\hline \hline
\end{tabular}
\end{center}
\caption{
Predicted masses (in MeV) of the $X(3872)$ HQSS partners
for two different values of the gaussian cutoff.
We use as input 3871.6 MeV and 3917.4 MeV for the $X(3872)$ and $X(3915)$
masses, respectively.
From this we find that the value of the ${\rm LO}$ couplings are
$C_{0a}= -3.53$ fm$^2$ and $C_{0b}= 1.59$ fm$^2$
for $\Lambda = 0.5\,{\rm GeV}$
and $C_{0a}= -1.06$ fm$^2$ and $C_{0b}= 0.27$ fm$^2$ for $\Lambda =1\,{\rm GeV}$.
Errors in our predicted masses are obtained by varying the strength of
the contact interaction in each channel by $\mp 15\%$,
which corresponds to the expected violations of HQSS
for the charm quark mass. 
} 
\label{tab:hqs-partners}
\end{table*}

We start by considering the $1^{++}$ $\rm D\bar{D}^*$ state, the $X(3872)$,
within the EFT formalism described in the previous section.
From HQSS we expect that the heavy meson-antimeson interaction in the
$1^{++}$ $\rm D\bar{D}^*$ and $2^{++}$ $\rm D^*\bar{D}^*$ channels
will be identical.
In terms of the EFT potential we have
\begin{eqnarray}
V^{\rm LO}(1^{++}) = V^{\rm LO}(2^{++}) \, , 
\end{eqnarray}
as deduced from Eqs.~(\ref{eq:contact2-a}-\ref{eq:contact2-b}).
From this, we automatically anticipate the existence of a isoscalar
$2^{++}$ $\rm D^*\bar{D}^*$ bound state with a binding energy
similar to that of the $X(3872)$.

To pinpoint the exact location of the $2^{++}$ partner of the $X(3872)$
we begin by determining the counterterm combination $C_{0a} + C_{0b}$.
We will use a gaussian regulator $f(x) = e^{-x^2}$ and the cut-off
values $\Lambda = 0.5\,{\rm GeV}$ and $1\,{\rm GeV}$.
As we are working in the isospin symmetric limit~\footnote{In what follows,
we use the isospin averaged masses $m_D=1867.2\,{\rm MeV}$ and
$m_D^*=2008.6\,{\rm MeV}$.}, we can consider the $X(3872)$
to be a $\rm D^*\bar{D}$ bound state with a binding
energy of $B_X(1^{++}) \simeq 4.2\,{\rm MeV}$,
from which we obtain $C_{0a} + C_{0b} = -1.94\,{\rm fm}^2$ ($-0.79\,{\rm fm}^2$)
for $\Lambda = 0.5\,{\rm GeV}$ ($1\,{\rm GeV}$).
Now we can predict that the mass of the $2^{++}$ state lies in the vicinity
of $4012\,{\rm MeV}$ (a value rather independent of the cut-off),
corresponding to a binding energy of $5\,{\rm MeV}$.
We call this state the $X(4012)$, and stress that this prediction is
independent of any assumptions about the molecular nature of
any other XYZ states, relying on HQSS alone.
We mention in passing that isospin breaking effects, even though crucial
for understanding certain decay properties of the $X(3872)$
(see, for example, Refs.~\cite{Gamermann:2009fv,Gamermann:2009uq}),
has not an appreciable effect in the spectroscopy problem~\footnote{
If we take into account isospin breaking within the formalism of
Ref.~\cite{Gamermann:2009uq}, we find that the position of
the $X(4012)$ moves by about $\sim 1\,{\rm MeV}$,
to $4013\,{\rm MeV}$. In this calculation we have ignored
the $I=1$ counterterm in the $J^{PC} = 1^{++}$ channel,
which (in absence of additional information) is considered
to be of order $Q^0$ and hence subleading.
}.

For predicting states beyond the $X(4012)$, we have to identify a particular
XYZ state as a further molecular partner of the $X(3872)$. 
In this way we will be able to determine the two contact range couplings
($C_{0a}$ and $C_{0b}$) and obtain the full spectrum of molecular states.
Two interesting candidates are the $X(3915)$~\cite{Uehara:2009tx} and
the $X(3940)$~\cite{Abe:2007jn,Abe:2007sya},
from which the first is the most promising.
The $X(3915)$ has been theorized to be a $0^{++}$ or $2^{++}$
$\rm D^* \bar D^*$ molecule in Refs.~\cite{Liu:2009ei,Branz:2009yt}.
Even though the work of Refs.~\cite{Liu:2009ei,Branz:2009yt}
cannot discriminate between the $0^{++}$ or $2^{++}$ quantum numbers~\footnote
{Branz et al.~\cite{Branz:2009yt} notice that the (little) known decay
properties of the $X(3915)$ are compatible with both assignments},
HQSS suggests that the most probable $J^{PC}$ value is $0^{++}$
instead of $2^{++}$, as the later would imply
a remarkable violation of HQSS.
On the contrary, the accommodation of the $X(3940)$~\cite{Abe:2007jn,Abe:2007sya}
within the HQSS pattern of molecular states faces a problem.
The $X(3940)$ decays strongly to $\rm D{\bar D^{*}}$, a feature compatible
with the expected properties for a $1^{+-}$ $\rm D^*{\bar D^{*}}$ axial state.
However, the production mechanism for the $X(3940)$ is more compatible
with a positive C-parity state than with a negative C-parity one:
this state is produced in the reaction $e^{+} e^{-} \to J / \Psi X(3940)$,
most probably via an intermediate virtual photon
($e^{+} e^{-} \to \gamma^{*} \to J / \Psi X$),
suggesting that the C-parity is positive.
Nonetheless this is not a definitive conclusion, and it may happen that
the $X(3940)$ resonance is being produced via two virtual photons,
see for example Refs.~\cite{Bodwin:2002fk,Bodwin:2002kk}
for a case in which this process is less suppressed
than naively expected.

At this point we notice that the identification of the $X(3915)$ as a 
$\rm D^*\bar{D}^*$ molecular state, though more promising than that
of the $X(3940)$, is not free of problems either.
In particular, the binding energy of the corresponding heavy vector
meson-antimeson system is of the order of $\sim 100\,{\rm MeV}$, 
which translates into a wave number of $\sim 450\,{\rm MeV}$.
This means that the $X(3915)$ lies not too far away from the limits
of what can be described within the EFT.
Its wave number indicates that a description in terms of mesons alone
may be incomplete and that the explicit inclusion of shorter range
components (e.g. tetraquark or charmonium-like) may be
necessary~\footnote{This is based on the observation that the mean quadratic
separation of the mesons is $\sqrt{\langle r^2 \rangle} = 0.5-0.8\,{\rm fm}$ 
depending on the cut-off.}.
However, the EFT framework is very helpful and convenient in this regard.
Abusing the limits of the EFT translates into a noticeable cut-off dependence
and a lack of convergence of the EFT expansion, that is, subleading
order corrections will be able to completely alter
the ${\rm LO}$ results.
As we will explain in the next paragraph, the cut-off dependence is numerically
small, and as we will check in the next section, the subleading order
corrections are moderate, but nonetheless under control.
All this indicates that the $X(3915)$ is probably more amenable to
an EFT treatment than naively expected.

From the assumption that the $X(3915)$ is a $0^{++}$ molecule
we can determine the counterterm combination $C_{0a} - 2\,C_{0b}$
and consequently the location of the six possible HQSS partners
of the $X(3872)$. 
The masses of the molecular states resulting from the previous identification
can be consulted in Table~\ref{tab:hqs-partners}.
We obtain a $0^{++}$ $\rm D\bar D$, $1^{+-}$ $\rm D\bar D^*$ and 
$1^{+-}$ $\rm D^*\bar D^*$ state which we call $X(3710)$,
$X(3820)$ and $X(3955)$.
The errors in Table~\ref{tab:hqs-partners} refer to uncertainties
owing to violations of HQSS in the charm sector (see Section~\ref{subsec:mQ}
for a detailed explanation).
The location of these three hadronic molecules is rather independent
of the cut-off.
If (instead of a gaussian regulator) we use a sharp cut-off,
there are small variations of about $\sim 1\,{\rm MeV}$
in the location of the states.
Curiously, the $X(3955)$ state we obtain is not far away
from the aforementioned $X(3940)$ molecular candidate.
A possible identification is suggestive but contingent on the eventual
determination of the quantum numbers (especially the C-parity) of the $X(3940)$.

\begin{table*}
\begin{center}
\begin{tabular}{|c|c|c|c|c|c|}
\hline \hline
$J^{PC}$ & $\rm H\bar{H}$ & $^{2S+1}L_J$ 
& $E$ $(\Lambda = 0.5\,{\rm GeV})$ & $E$ $(\Lambda = 1\,{\rm GeV})$ & ${\rm
  Exp}$~\cite{Nakamura:2010zzi}  \\
\hline
$0^{++}$ & $ D\bar{D}$ & $^1S_0$
& $3708$ & $3720$ & $-$ \\
\hline
$1^{++}$ & $ D^*\bar{D}$ & $^3S_1$-$^3D_1$ 
& Input & Input  & $3872$ \\
$1^{+-}$ & $ D^*\bar{D}$ & $^3S_1$-$^3D_1$ 
& $3816$ & $3823$ & $-$ \\
\hline
$0^{++}$ & $ D^*\bar{D}^*$ & $^1S_0$-$^5D_2$  
& Input & Input & $3917$ \\
$1^{+-}$ & $ D^*\bar{D}^*$ & $^3S_1$-$^3D_1$  
& $3954$ & $3958$ & $3942$ \\
$2^{++}$ & $ D^*\bar{D}^*$ & $^1D_2$-$^5S_2$-$^5D_2$-$^5G_2$ 
& $4015$ & $4014$ & $-$ \\
\hline \hline
\end{tabular}
\end{center}
\caption{
Predicted masses (in MeV) of the $X(3872)$ HQSS partners
 when the  OPE potential is included. We display results 
for two different values of the gaussian cutoff.  Now, we find
$C_{0a}= -3.46$ fm$^2$ and $C_{0b}= 1.98$ fm$^2$, and $C_{0a}= -0.98$
fm$^2$ and $C_{0b}= 0.69$ fm$^2$, for $\Lambda=0.5$ and 1 GeV,
respectively.} \label{tab:hqs-partners-OPE-a}
\end{table*}

\section{Subleading Order Corrections}
\label{sec:subleading}

In this section we explore the impact of
the main subleading order contributions.
For this we must take into account the existence of two different,
unrelated expansions in the EFT formalism we are proposing.
The first is the expansion in terms of inverse powers of the heavy quark mass
and the second the standard power counting expansion.
In the previous section we have assumed exact, instead of approximate, HQSS.
The existence of $1/m_Q$ deviations from the heavy quark limit implies
that the location of the molecular partners of the $X(3872)$
will be subjected to uncertainties.
Apart from this, the power counting expansion indicates that we should
take into account two important subleading order corrections appearing
at order $Q^0$ and $Q^1$ respectively: 
the OPE potential and the particle coupled channel effects.
As we will see, the induced shifts in the position of the molecular states
from these subleading interactions will in general agree with
(and in the case of the OPE potential be smaller than)
the {\it a priori} EFT expectations,
thus confirming the robustness of the molecular spectrum
we have deduced so far.

\subsection{The $1/m_Q$ Corrections}
\label{subsec:mQ}

The ${\rm LO}$ potentials for the $1^{++}$ and $2^{++}$ heavy meson molecules
-- the $X(3872)$ and the theorized $X(4012)$ -- are only identical
in the heavy quark limit.
Thus the existence of the $X(4012)$ may be contingent
on the size of HQSS violations stemming
from the finite charm quark mass.
In general, we expect the heavy meson-antimeson potentials
of Eqs.~(\ref{eq:contact2-a}-\ref{eq:contact2-b}) and
(\ref{eq:coni}-\ref{eq:conf})
to deviate from the heavy quark limit
by a quantity of the order of
\begin{eqnarray}
V^{\rm LO}_{(m_Q = m_c)} = V^{\rm LO}_{(m_Q = \infty)} \,
\left( 1 +{ \cal O}(\frac{\Lambda_{\rm QCD}}{m_c}) \right) \, ,
\end{eqnarray}
where $m_c$ is the charm quark mass ($\sim 1.5\,{\rm GeV}$) and
$\Lambda_{\rm QCD} \sim 200\,{\rm MeV}$, translating into an expected $15\%$
violation of HQSS for the ${\rm LO}$ contact range potentials.
The exceptions are the potentials of the $1^{++}$ ${\rm D\bar D^*}$ and
$0^{++}$ ${\rm D^*\bar D^*}$ channels, as they are fixed
to reproduce the position of the $X(3872)$ and $X(3915)$ states.

Assuming this $15\%$ uncertainty in the contact range interactions,
we obtain the error bars of Table~\ref{tab:hqs-partners}.
As can be seen, the prediction of the $2^{++}$ partner of the $X(3872)$
is robust with respect to this theoretical error source.
Other molecular states show moderate uncertainties of the order of
$10-20\,{\rm MeV}$ in the binding energies.
As we will see in the following subsections,
these uncertainties are a bit smaller (but yet of the same order)
as the subleading order corrections coming from pion exchanges
and the particle coupled channels.
   
\subsection{One Pion Exchange}

A possible issue with the present EFT treatment of the $X(3915)$
as a $\rm D^*\bar{D}^*$ molecule is whether the OPE potential
is really perturbative in this case.
With a wave number of $\gamma \simeq 450\,{\rm MeV}$, the wave function
of the $X(3915)$ probes the intermediate distances where the tensor
component of the OPE potential is stronger than the central one.
In this regard and according to the arguments in Ref.~\cite{Valderrama:2012},
we expect tensor OPE to become non-perturbative at a critical
center-of-mass momentum of $k_{\rm crit} \simeq 420\,{\rm MeV}$
for the $0^{++}$ $\rm D^*\bar{D}^*$ channel~\footnote{As we will
comment later, this number is subjected to large uncertainties.}.
Consequently the $X(3915)$ lies at the edge of the domain of validity of
the EFT description we are using, in which pion exchanges are
perturbative (i.e. small), and may require a more sophisticated EFT
with non-perturbative pions.
If we were predicting the binding energy of this molecular state,
the previous observation would translate into a large uncertainty
in the calculations of the order of $100\%$.
The cause of this uncertainty will be high energy fluctuations 
in the meson-antimeson loops generated by the pion exchanges.
However, as the binding energy is used as the input of the calculation,
the uncertainty can manifest either in the value of the $C_{0a}$ and
$C_{0b}$ counterterms, thus subjecting the predictions
of Table \ref{tab:hqs-partners} to large uncertainties,
or in a failure of the theory to converge once
subleading order corrections are included.

Yet there are two mitigating circumstances that may increase
the expected breakdown scale of the present EFT calculation,
thus turning the predictions much more reliable.
On the one side, the critical momenta above which the tensor OPE
is no longer perturbative~\cite{Valderrama:2012}
are subjected to considerable uncertainties stemming from
the far-from-perfect separation of scales in heavy meson molecules.
In particular the value of the critical momenta may be up to $50\%$ larger
than expected, in which case the $X(3915)$ will lie well within
the range of applicability of the EFT with perturbative pions.
On the other, we are limiting ourselves to the spectroscopy problem.
It is worth noticing in this respect that the spin-spin structure of
the $C_{0b}$ contact operator is very similar to the one we obtain
from pseudoscalar meson exchange for the non-tensor piece
of the interaction.
We thus expect the $C_{0b}$ operator to be able to partially absorb the shift
in the binding energy generated by the pion exchanges.
However, the effect is restricted to the central component of OPE,
which is much better behaved than the tensor one
at short distances.
Of course, the definitive test is to recalculate the position
of the predicted molecular states after the non-perturbative
inclusion of the OPE potential.

The distinctive feature of the OPE potential is that it can mix
different partial waves.
For the non-perturbative calculation we follow the formalism of
Ref.~\cite{Nieves:2011zz},
where we considered the partial wave decomposition of tensor OPE in detail
for the $\rm P\bar{P}^*$ particle channel.
The extension to the $\rm P^*\bar{P}^*$ case is trivial and only
entails a change in the partial waves that are actually
coupled, which can be consulted in Table~\ref{tab:hqs-partners-OPE-a}.
For determining the value of the $C_{0a}$ and $C_{0b}$ counterterms
we fix the binding energies of the $X(3872)$ and $X(3915)$
respectively.
This procedure generates again a total of six molecular states,
four of them predictions,
as can be seen in Table \ref{tab:hqs-partners-OPE-a}.
We can appreciate that the binding energies of the states are relatively
stable with respect to the iteration of the OPE potential.
The most affected state is the $0^{++}$ $\rm D\bar{D}$ molecule,
the $X(3710)$ state, for which the binding energy can increase
by almost $10\,{\rm MeV}$ for $\Lambda = 1\,{\rm GeV}$
with respect to the pionless case.
Curiously, for this state there is no OPE contribution to the finite range
potential: two pseudoscalar objects cannot exchange a single pion,
so the first non-trivial contribution to the finite range potential
comes from two pion exchange, which we have not considered here. 
The dynamics of this state is solely controlled by
the $C_{0a}$ counterterm.
In this regard, when the contact range potential is adjusted to reproduce
the binding energies of the $X(3872)$ and $X(3915)$ states
in presence of the OPE potential,
the change in the value of the $C_{0a}$ coupling
is not counterbalanced by a pion exchange contribution in
the $0^{++}$ $\rm D\bar{D}$ case.  
Hence, we find a larger shift in the location of this molecular state
in comparison to the others.

Indeed we find relatively small shifts in the energy of states other
than the $X(3710)$
(see Tables \ref{tab:hqs-partners} and \ref{tab:hqs-partners-OPE-a}). 
The aforementioned observation seem to confirm the suspicions
about the role of the $C_{0b}$ operator
which could effectively accommodate most of the OPE effects. 
Equivalently, the inclusion of the OPE potential produces a larger change
in the $C_{0b}$ coupling than in the $C_{0a}$ one,
and actually we find that the bulk of the change in $C_{0b}$ is given
by the size of the contact piece of OPE
\begin{eqnarray}
C_{0b} \to C_{0b} - \frac{g^2}{2 f_{\pi}^2} \, ,
\end{eqnarray} 
where $g^2/2f_\pi^2 \sim 0.4\,{\rm fm}^2$, see Appendix \ref{app:LO-pot}
for details

The natural conclusion of the previous calculations is that
OPE is indeed perturbative in heavy meson molecules.
There is however a caveat: if the spectroscopy problem is rather
insensitive to the OPE potential, how can we appreciate whether
OPE is perturbative or not in the molecular states?
A partial answer is that $C_{0b}$ is only expected to absorb effects
coming from central OPE, but not from tensor OPE,
which is the problematic piece.
But there is another answer, that lies in the observation that
non-perturbative OPE would entail a significant change
in the power counting of the contact range operators.
As happens in nuclear EFT, the ${\rm LO}$ counterterms that are able
to renormalize the scattering amplitude with perturbative
pions~\cite{Kaplan:1998tg,Kaplan:1998we} are not enough
to renormalize the corresponding non-perturbative
formulation~\cite{Nogga:2005hy}.
In particular, for heavy meson EFT we have that while perturbative OPE
requires two counterterms ($C_{0a}$ and $C_{0b}$),
for non-perturbative OPE this number
is at least five~\cite{Valderrama:2012}.
As the shifts in the binding energies are only weakly cut-off dependent,
we do not require new counterterms and we can confidently conclude
that OPE is perturbative.

We mention in passing that an alternative possibility to check whether OPE
is perturbative is the description of bound state properties
that depend on the existence of a $D$-wave component of the wave function,
which is a typical signature of the tensor force. 
It turns out that $D$-wave probabilities of the molecular states
are quite small (from $1-4\%$, with a strong cut-off dependence~\footnote{
One should keep in mind that the $D$-wave probability is not {\it per se}
an observable quantity, meaning that we should not be worried
by its moderate cut-off dependence.}).
Unfortunately, it looks difficult to find experimental observations
that could depend on (and consequently constraint) the $D$-wave
components of the molecular wave-functions.
Nevertheless, the small size of the $D$-wave probabilities is consistent
with the expectation of them to be a second order correction
in perturbation theory.

\subsection{Particle Coupled Channels}
\label{sub:particle-CC}

\begin{table*}
\begin{center}
\begin{tabular}{|c|c|c|c|c|}
\hline \hline
$J^{PC}$ & $\rm H\bar{H}$ & 
$E - i \Gamma/2$ $(\Lambda = 0.5$ GeV)
& $E - i \Gamma/2$ $(\Lambda = 1$ GeV) & ${\rm Exp}$~\cite{Nakamura:2010zzi} \\
\hline
$0^{++}$ & $ D\bar{D} , D^*\bar{D}^*$ 
& $3658$ & $3669$ &  $-$ \\
\hline
$1^{++}$ & $ D^*\bar{D}$
& Input & Input  & $3872$ \\
$1^{+-}$ & $ D^*\bar{D}, D^*\bar{D}^*$ 
& $3730$ & $3739$ & $-$ \\
\hline
$0^{++}$ & $ D\bar{D}, D^*\bar{D}^*$ 
& $3917 - \frac{i}{2}\,23$ & $3917 - \frac{i}{2}\,50$ & $3917 \pm 3- \frac{i}{2}\,28^{+10}_{-9}$  \\
$1^{+-}$ & $ D^*\bar{D}, D^*\bar{D}^*$ 
& $3979 - \frac{i}{2}\,24$ & $3979 - \frac{i}{2}\,39$ & $3942\pm 9 - \frac{i}{2}\,37^{+27}_{-17}$ \\
$2^{++}$ & $ D^*\bar{D}^*$ 
& $4012$ & $4012$  & $-$ \\
\hline \hline
\end{tabular}
\end{center}
\caption{
Predicted masses and widths (in MeV) of the $X(3872)$ HQSS partners 
when coupled channels effects are included. The contact terms are
adjusted to reproduce the $X(3872)$ and $X(3915)$ masses, while OPE
effects are neglected.  We find
$C_{0a}= -4.16$ fm$^2$ and $C_{0b}= 2.21$ fm$^2$, and $C_{0a}= -1.14$
fm$^2$ and $C_{0b}= 0.35$ fm$^2$, for $\Lambda=0.5$ and 1 GeV,
respectively.
} \label{tab:hqs-partners-CC-a}
\end{table*}

\begin{table*}
\begin{center}
\begin{tabular}{|c|c|c|c|c|}
\hline \hline
$J^{PC}$ & $\rm H\bar{H}$ & 
$E - i \Gamma/2$ $(\Lambda = 0.5$ GeV)
& $E - i \Gamma/2$ $(\Lambda = 1$ GeV) & ${\rm Exp}$~\cite{Nakamura:2010zzi} \\
\hline
$0^{++}$ & $ D\bar{D} , D^*\bar{D}^*$ 
& $3690$ & $3694$ &  $-$ \\
\hline
$1^{++}$ & $ D^*\bar{D}$
& Input & Input  & $3872$ \\
$1^{+-}$ & $ D^*\bar{D}, D^*\bar{D}^*$ 
& $3782$ & $3782$ & $-$ \\
\hline
$0^{++}$ & $ D\bar{D}, D^*\bar{D}^*$ 
& $3939 - \frac{i}{2}\,12$ & $3937 - \frac{i}{2}\,31$ & $3917 \pm 3- \frac{i}{2}\,28^{+10}_{-9}$  \\
$1^{+-}$ & $ D^*\bar{D}, D^*\bar{D}^*$ 
& $3984 - \frac{i}{2}\,17$ & $3982 - \frac{i}{2}\,29$ & $3942\pm 9 - \frac{i}{2}\,37^{+27}_{-17}$ \\
$2^{++}$ & $ D^*\bar{D}^*$ 
& $4012$ & $4012$  & $-$ \\
\hline \hline
\end{tabular}
\end{center}
\caption{
Predicted masses and widths (in MeV) of the $X(3872)$ HQSS partners 
when coupled channels effects are included. The contact terms are
fixed to the values given in the caption of
Table~\ref{tab:hqs-partners} (i.e.,  
they are adjusted to reproduce the $X(3872)$ and $X(3915)$ masses
neglecting coupled channel effects). Moreover, OPE interactions are not taken
into account either.
}
 \label{tab:hqs-partners-CC-b}
\end{table*}

As previously discussed particle coupled channel effects are suppressed
by two orders in the EFT expansion.
From this we expect coupled channel effects in the binding energies of
the $0^{++}$ and $1^{+-}$ states to scale as 
\begin{eqnarray}
| \Delta E_B | \simeq | E_B |\,
{\left( \frac{\gamma_B}{\Lambda_C} \right)}^2 \, , \label{eq:expect}
\end{eqnarray}
where $E_B$ is the binding energy, $\gamma_B = \sqrt{-2 \mu E_B}$
the wave number of the bound state and $\Lambda_C$ the typical
momentum scale of the coupled channel under consideration,
which we consider to be a hard scale
$\Lambda_C \sim \Lambda_0$.
The estimation above translates into an uncertainty of
around $30\,{\rm MeV}$ ($40\,{\rm MeV}$) for the $0^{++}$
($1^{+-}$) coupled channel, where we have employed
the wave number of the deepest bound state within
the coupled channel.
It should be noticed that in the case of the $\rm D^*\bar{D}^*$ molecular
states the energy shift is complex, as the $0^{++}$ ($1^{+-}$) state
can decay into a $\rm D\bar{D}$ ($\rm D^*\bar{D}$)
meson-antimeson pair.

In contrast to the OPE corrections, the counterterm structure stemming
from HQSS is not expected to be able to absorb the kind of divergences
associated with the coupled channel calculations,
meaning that the actual error in the calculation
will probably saturate the previous bound.
We can check the EFT {\it a priori} estimates given above by
means of a concrete calculation in which the particle coupled channel
effects are fully taken into account.
However, as we will see, this task is not trivial, specifically in
what regards to the choice of the appropriate regularization scheme.
To illustrate this point we can study the perturbative estimate of the
binding energy shift induced by  coupled channel dynamics.
If we consider a small change in the potential
\begin{eqnarray}
V \to V + \delta V \, ,
\end{eqnarray} 
the perturbative correction to the binding energy is expected to be
\begin{eqnarray}
\delta E = \langle \Psi | \delta V | \Psi \rangle \, , 
\end{eqnarray}
where $| \Psi \rangle$ is the wave function of the bound state.
In the case of particle coupled channels, the $\delta V$ operator reads
\begin{eqnarray}
\delta V_{\alpha} = V_{\alpha \beta} G_{0,\beta}(E) V_{\beta \alpha} \, ,
\end{eqnarray}
where $\alpha$ represents the channel we are interested in,
$\beta \neq \alpha$ the other channel and $V_{\alpha \beta}$
the transition potential from channel $\alpha$ to $\beta$,
which is proportional to the $C_{0b}$ contact operator.
We can distinguish two cases, depending on whether the unperturbed
energy $E$ is (a) above or (b) below the $\beta$ channel threshold.
The most interesting case is (a), corresponding to the modification
of the $0^{++}$ or $1^{+-}$ $\rm D^*\bar{D}^*$ molecular state energies
($\alpha$ channel) by the $0^{++}$ $\rm D\bar D$ or
$1^{+-}$ $\rm D\bar{D}^*$ states ($\beta$ channel),
which lie in the continuum for the energies relevant
for the $\alpha-$channel states.
A direct calculation yields
\begin{eqnarray}
\delta E_{\alpha} &=& {\left[
C_{\alpha \beta}(\Lambda) \, \int_{\Lambda}
\frac{d^3\vec{p}}{(2 \pi)^3}\,\Psi_{\alpha}(\vec{p}\,) \right]}^2 
\nonumber \\
&\times& \int_{\Lambda} \frac{d^3\,\vec{q}}{(2 \pi)^3}\,
\frac{2\mu}{k_{\beta}^2 - \vec{q}^{\,2} + i \epsilon} \, ,
\label{ed:Delta-E-cc}
\end{eqnarray}
where $\Psi_{\alpha}(\vec{p}\,)$ is the wave function 
of the $\alpha$ bound state,
$k_{\beta}^2 = - \gamma_{\alpha}^2 + \Lambda_{C}^2$ the momentum
of the heavy meson-antimeson pair above the threshold,
$\gamma_{\alpha}$ the wave number of the $\alpha$ bound state
below threshold and $\Lambda_{C}$ the coupled channel
momentum scale.
The integrals are assumed to be regularized with a cut-off $\Lambda$
and an arbitrary regulator function that we have not specified yet.
The $C_{\alpha \beta}$ transition contact operator can be identified
with $2\,C_{0b}$ ($\sqrt{3}\,C_{0b}$) in the $0^{++}$ ($1^{+-}$)
molecular state.

As can be seen, if $\Lambda < k_{\beta}$, a natural thing to expect
if the momentum separation of the coupled channels is a hard scale,
the perturbative correction to the binding energy is effectively
suppressed by a factor of $k_{\beta}^2 \sim \Lambda_{C}^2$ :
\begin{eqnarray}
\delta E_{\alpha} &=& \frac{2\mu}{k_{\beta}^2}\,
{\left[ C_{\alpha \beta}(\Lambda) \, \int_{\Lambda}
\frac{d^3\vec{p}}{(2 \pi)^3}\,\Psi_{\alpha}(\vec{p}\,) \right]}^2 
\nonumber \\
&\times& \int_{\Lambda} \frac{d^3\,\vec{q}}{(2 \pi)^3}\,
\left( 1 + \frac{\vec{q}^{\,2}}{k_{\beta}^2} + 
\frac{\vec{q}^{\,4}}{k_{\beta}^4} + \dots \right)
\, ,
\label{ed:Delta-E-cc-exp}
\end{eqnarray}
that is, by two powers in the counting.
However, we can also appreciate that the $\delta E_{\alpha}$ correction
is strongly scale dependent.
On the one hand, if we consider that the wave function behaves as
\begin{eqnarray}
\Psi_{\alpha}(\vec{p}\,) \propto \frac{1}{\vec{p}^{\,2} + \gamma_{\alpha}^2} \, ,
\end{eqnarray}
we see that the integral in the first line of the expression for
the energy shift of Eq.~(\ref{ed:Delta-E-cc})
diverges as $\Lambda$:
\begin{eqnarray}
\int_{\Lambda} \frac{d^3\vec{p}}{(2 \pi)^3}\,\Psi_{\alpha}(\vec{p}\,)
\propto \Lambda \, .
\end{eqnarray}
On the other, the integral related to the decay into the continuum state
$\beta$ in the second line of Eq.~(\ref{ed:Delta-E-cc})
diverges as $\Lambda$.
However, had we re-expanded the propagator $1 / (k_{\beta}^2 - \vec{q}^{\,2})$
in inverse powers of $k_{\beta}$ (as mandated by power counting), 
the power-law divergence would have worsened to $\Lambda^3$,
see Eq.~(\ref{ed:Delta-E-cc-exp}).
Putting all the pieces together (and ignoring the propagator re-expansion),
the total divergence in the energy shift is given
by $C_{\alpha \beta}^2\,\Lambda^3$.

This divergent behaviour tell us that we are required to include
counterterms to absorb them~\footnote{It is important to comment at this point
that the different divergences we are discussing correspond to particular
choices of how to expand in terms of power counting.
The full, non-perturbative coupled channel calculation is free of divergences.}.
The problem is that the counterterms renormalizing
the coupled channel dynamics are higher order.
In principle the $C_{0b}$ operator could do the job,
but we need to take into account that this contact operator is already
determined by the condition of reproducing the binding energy
of a molecular state.
Thus, we do not expect $C_{0b}$ to balance for the particle
coupled channel effects.
The renormalization group behaviour of $C_{0b}$ is approximately given by 
\begin{eqnarray}
C_{0b}(\Lambda) \propto \frac{1}{\mu\,\Lambda} \, ,
\end{eqnarray}
for large $\Lambda$, see Eq.~(\ref{eq:C0-running}).
This means in turn that $C_{0b}$ can absorb the piece proportional
to $\Lambda^2$ of the coupled channel divergence,
\begin{eqnarray}
{\left[ C_{\alpha \beta}(\Lambda) \, \int_{\Lambda}
\frac{d^3\vec{p}}{(2 \pi)^3}\,\Psi_{\alpha}(\vec{p}\,) \right]}^2 
\propto \Lambda^0 \, ,
\end{eqnarray} 
since $C_{\alpha \beta}$ is proportional to $C_{0b}$.
Thus, we are left with a residual cut-off dependence of $\Lambda$
in the best case.
The worst case scenario is however when the cut-off and the coupled channel
scale coincide, $\Lambda \sim \Lambda_{C}$, in which case
the $\beta$-channel integral peaks.
This can be easily appreciated if we use a sharp cut-off regulator
\begin{eqnarray}
2 \pi^2 \, \int \frac{d^3\vec{q}}{(2 \pi)^3}\,
\frac{\theta(\Lambda - |\vec{q}\,|)}{k_{\beta}^2 - \vec{q}^{\,2} + i \epsilon}
&=& - \left[
\Lambda + i \frac{\pi }{2} k_{\beta} \, \theta(\Lambda - k_{\beta})
\right] \nonumber \\
&+& \frac{k_{\beta}}{2}\, \log{\left| \frac{\Lambda - k_{\beta}}{\Lambda + k_{\beta}} \right|} \, ,
\label{eq:CC-sharp}
\end{eqnarray}
where the real part of the integral diverges at
$\Lambda = k_{\beta} \sim \Lambda_C$,
a very puzzling situation (see a related discussion in
Ref.~\cite{Khemchandani:2011mf}).
Of course, the problem can be solved by using a smoother regulator,
as the gaussian scheme that we have been employing along this work.
In this case the real part integral will show a maximum (but not diverge)
at $\Lambda \sim \Lambda_C$.
This signals the transition from a power counting in which $\Lambda_C$
is a hard scale to a different one in which it is a soft scale.

All this indicates that one should presumably add new counterterms
at ${\cal O}(Q)$ to soften the cutoff dependence and
make the EFT renormalizable again.
At this point it is worth mentioning that the EFT treatment of coupled channel
dynamics has been only discussed for the case
in which $\Lambda_C$ is a soft scale~\cite{Cohen:2004kf,Lensky:2011he}.
However, the corresponding analysis for the $\Lambda_C \sim \Lambda_0$ case
has not been done yet and will be left for future research~\cite{Nieves:2012}.
Independently of the exact form of the power counting for coupled channel
dynamics, it is clear that higher orders will introduce new unknown constants
that cannot be fixed at the moment owing to
the scarce experimental data available.
Nevertheless, here we will present full non-perturbartive results
including coupled channel effects.
Even though the energy shifts thus obtained will be cutoff (and regulator)
dependent, they will not vastly deviate from the {\it a priori} estimates
of Eq.~(\ref{eq:expect}), reinforcing the (qualitative) reliability
of the ${\rm LO}$ predictions.
Of course, had we included all the relevant ${\cal O}(Q)$ counterterms
(and known the entire experimental spectrum of ${\rm D^{(*)} \bar D^{(*)}}$
states), the deviations would have decreased.
Furthermore, we notice that coupled channel effects produce changes
in $C_{0a}$ and $C_{0b}$ comparable in magnitude to those
we should expect from violations of HQSS, that is,
about $15\%$ for the charm quark mass. 

The non-perturbative calculation of the coupled channel effects is
presented in Table \ref{tab:hqs-partners-CC-a}.
As in previous cases, we have adjusted the $C_{0a}$ and $C_{0b}$ counterterms
to reproduce the $X(3872)$ and $X(3915)$ masses.
We have searched for the poles of the scattering amplitude in the first
and second Riemann sheets (we refer to Ref.~\cite{Nieves:2001wt} for
further details on this subject).
The former are to be interpreted as bound states, while the later correspond
to the location of resonant states in the complex plane,
where the real part of the pole position is the mass and
the imaginary part is half the decay width
($E_{\rm pole} = M - \frac{i}{2}\,\Gamma$).
As can be seen in Table \ref{tab:hqs-partners-CC-a},
the location of the $1^{+-}$ $\rm D\bar{D}^*$ and $\rm D^*\bar{D}^*$ states
have been shifted by about $80-85\,{\rm MeV}$ and $25\,{\rm MeV}$ respectively.
The correction to the binding energy of the $\rm D\bar{D}^*$ state is large,
saturating and even exceeding the EFT expectation.
In the case of the $0^{++}$ $\rm D\bar{D}$ partner of the $X(3915)$,
the shift in the position of the state is of $40-50\,{\rm MeV}$,
of the order of the EFT expectation. 
Had we use a sharp cut-off instead of a gaussian one,
the location of the bound and resonant states would have drastically changed
for $\Lambda=0.5\,{\rm GeV}$.
This is not surprising in view of Eq.~(\ref{eq:CC-sharp})
and the related discussion.
However, for larger cut-offs such as $\Lambda=1\,{\rm GeV}$,
variations are significantly smaller and
the results are similar to those obtained in the gaussian cut-off scheme.

To further check the uncertainties affecting our results, 
we have also considered the alternative option of using
the counterterm values of the original uncoupled calculation
to estimate the coupled channel effects,
in which case we obtain the results of Table~\ref{tab:hqs-partners-CC-b}.
In this second scheme we observe that the change in the position of
the states agrees much better with the EFT expectations:
the energies of the $1^{+-}$ $\rm D\bar{D}^*$ and $\rm D^*\bar{D}^*$ states
consistently change by about $35\,{\rm MeV}$,
while in the $0^{++}$ $\rm D\bar{D}$ and $\rm D^*\bar{D}^*$ states we end up
with an energy shift of about $20-25\,{\rm MeV}$. 
The reason for the additional stabilization of the calculations may be
that we are not forcing the reproduction of the $X(3915)$
in a cut-off window in which we may not expected to obtain
this state (owing to the large coupled channel corrections).
In this case the $X(3915)$ state, which we do not adjust now, 
shifts is mass to around $3940 \,{\rm MeV}$,
with a width of about $15-30 {\rm MeV}$,
depending on the value of the cutoff. 
If we consider that the uncertainties coming from the $1/m_Q$ corrections are
of the order of $15-30\,{\rm MeV}$,
the properties of the $0^{++}$ $\rm D^*\bar{D}^*$ state could certainly
be accommodated with the existing experimental data for this resonance
($M = 3917 \pm 3$ and
$\Gamma = 28^{+10}_{-9}\,{\rm MeV}$~\cite{Nakamura:2010zzi}).

\section{HQSS and Decay Properties}
\label{sec:decay}

The dynamics of the molecular states studied in this work is solely determined,
within our approach, by the re-interaction of
the open charm channels $\rm D^{(*)}\bar D^{(*)}$. 
We have ignored hidden charm channels like, for example, the $J/\Psi\, \omega$
or $\eta_c \, \omega$. 
We expect these latter channels to have little effect on the inner structure
and masses of the molecular states, as suggested by explicit calculations
performed in Refs.~\cite{Gamermann:2006nm, Gamermann:2007fi,Molina:2009ct}. 
Yet, within the EFT approach, it has been customary to ignore hidden
charm channels in the study of the $X(3872)$ resonance,
see e.g. Ref. \cite{AlFiky:2005jd}.
Analogously, the hidden bottom channels have also been ignored
in the recent studies~\cite{Voloshin:2011qa, Mehen:2011yh}
of the $Z(10610)$ and $Z'(10650)$ molecular states in the bottom sector. 
Nevertheless, the hidden charm channels can play an important role
in the decay of some of the states described here, especially if
they are placed below the open charm $\rm D^{(*)}\bar D^{(*)}$
thresholds. 
Moreover, the $J/\Psi$ meson provides a clear experimental signature and
thus its decay modes are often used in the detection of the XYZ states. 
The detailed study of the hidden charm decays of the molecular states
described here is beyond the scope of this work and
we left it for future research. 

However, the generic decay properties of the molecular states
can be discussed at the qualitative level in the basis of HQSS. 
If we ignore phase space effects,
HQSS predicts~\cite{Voloshin:2011qa, Mehen:2011yh} for the total widths:
\begin{eqnarray}
\Gamma(1^{++}) &=& \Gamma(2^{++}) \nonumber \\ &=&
\frac{3}{2}\,\Gamma_{\rm D\bar{D}}(0^{++}) 
- \frac{1}{2}\,\Gamma_{\rm D^*\bar{D}^*}(0^{++}) \, ,  \label{eq:decay-a} \\
\Gamma_{\rm D\bar{D}^*}(1^{+-}) &=& \Gamma_{\rm D^*\bar{D}^*}(1^{+-}) \, ,
\label{eq:decay-b}
\end{eqnarray}
where we denote each molecular state by its quantum number $J^{PC}$ and
additionally its particle content if necessary.
As noticed in Ref.~\cite{Mehen:2011yh}, the relations above can also
be obtained within the EFT framework we advocate by promoting
the $C_{0a}$ and $C_{0b}$ couplings to complex values.
In this way, one can implicitly take into account the multiple decay channels
of the molecular states (as with an optical potential).
In contrast with the bottom sector, where the previous relationships
were derived, we expect however noticeable corrections
to Eqs.~(\ref{eq:decay-a}) and~(\ref{eq:decay-b})
in the charm sector.
The reason is that both HQSS violations and phase space corrections
are larger in the charm sector than in the bottom one.
The relations above involve total widths and do not necessarily hold
for decays into open charm channels, where phase space corrections
are crucial and indeed forbid some decays. 
For instance, if we pay attention to Eq.~(\ref{eq:decay-a}),
and since $\Gamma_{\rm D\bar{D}}(0^{++}) = 0$ for an open charm decay
into $\rm D^*\bar D^*$~\footnote{Note that in the infinitely heavy quark
limit the $\rm D$ and $\rm D^*$ mesons are degenerated. Thus,
the $\rm D^*\bar D^*$ decay channel could be open
depending on the binding energies.},
we will have to conclude that
$\Gamma(1^{++}) = \Gamma(2^{++})= \Gamma_{\rm D^*\bar{D}^*}(0^{++})=0 $. 
However, we find a partial decay width of the order of tens of MeV 
for the $X(3915)$ state into $D\bar{D}$.

Nonetheless, a clear implication of the relationships above is that 
the $X(4012)$ should be a relatively narrow state,
just like the $X(3872)$.
In addition, if we assume $\Gamma(1^{++})$ and $\Gamma(2^{++})$ to be much
smaller than the other decay widths, then we can estimate the total width
of the $X(3710)$ resonance to be a third of the $X(3915)$ width
($\Gamma_{\rm D^*\bar{D}^*}(0^{++}) = 28^{+10}_{-9}$ according
to the PDG~\cite{Nakamura:2010zzi}),
yielding $\Gamma_{\rm D\bar{D}}(0^{++}) \sim 10\,{\rm MeV}$.
For the $X(3815)$ and $X(3955)$ resonances the situation is similar:
HQSS without phase space considerations predicts them to have
the same width, but if one takes into account the large $\rm D^*\bar{D}$
contribution to the $X(3955)$,
the $X(3815)$ should be narrower than its partner.

\section{Discussion and Conclusions}
\label{sec:conclusion}

In this work we have argued that the application of HQSS to
the charmed meson-antimeson system,
combined with the identification of the $X(3872)$ and $X(3915)$ resonances
as isoscalar $\rm D \bar D^{*}$ and $\rm D^* \bar D^{*}$ molecules,
implies the existence of four molecular partners
of these two states (Table \ref{tab:hqs-partners}).
This prediction is subjected to a series of uncertainties,
namely the approximate nature of HQSS (especially for the charm sector),
the effect of the OPE potential and
the impact of the particle coupled channel dynamics.
We have estimated the size of these corrections within the EFT framework
and concluded that the HQSS pattern of molecular states is rather stable
(Tables \ref{tab:hqs-partners-OPE-a}, \ref{tab:hqs-partners-CC-a} and
\ref{tab:hqs-partners-CC-b}).
In contrast, the exact location of the molecular partners is subjected
to moderate uncertainties of up to $40-50\,{\rm MeV}$ for the most bound cases,
in agreement with the EFT expectations.

The determination of the $\rm D^{(*)}\bar D^{(*)}$ family of bound states
hinges on the assumption that the $X(3872)$ and the $X(3915)$ states
are molecular.
In this regard we find it worth commenting that,
while the identification of the $X(3872)$ as a $1^{++}$ loosely
bound $\rm D\bar{D}^*$ state is a widely accepted hypothesis,
the case for the molecular nature of the $X(3915)$ is less compelling
but nevertheless still compatible with the experimental information
available for this resonance.
Thus we expect the conclusions solely derived from the $X(3872)$
to be more solid and less speculative
than those depending on the $X(3915)$.

In this regard the tentative $2^{++}$ ${\rm D^* \bar{D}^* }$ partner
of the $X(3872)$, which we have called the $X(4012)$ in reference to its
predicted mass, see Table \ref{tab:hqs-partners}, is probably
the most robust and model independent prediction of
the present work.
The $X(4012)$ is not affected by particle coupled channel effects
and its mass only varies mildly, by about $2-3\,{\rm MeV}$,
when the OPE potential is included.
Perhaps in the real world the $1/m_Q$ effects may be larger
than we have estimated or there may be a further and unexpected
subleading correction that turns out to be large.
In this case the $X(4012)$ state might move slightly up above
the $\rm D^*\bar D^*$ threshold and become virtual
or might descend to a lower mass region.
Be as it may, we are quite confident about the existence of a molecular state
with these quantum numbers close to the $\rm D^*\bar D^*$ threshold.

The prediction of new $\rm D^{(*)}\bar D^{(*)}$ states beyond the $X(4012)$
requires the identification of a further XYZ state (besides the $X(3872)$)
as a charmed meson-antimeson molecule.
The $X(3915)$ is a good candidate, which we assume to be 
a $0^{++}$  $\rm D^*\bar D^*$ bound state.
Of course we notice that the  molecular interpretation of the $X(3915)$~\cite{Liu:2009ei,Branz:2009yt,Ding:2009vd},
while plausible, is not so well-established.
Consequently the three additional $0^{++}$ ${\rm D\bar{D}}$ and $1^{+-}$
${\rm D\bar{D}^*}$ and ${\rm D^*\bar{D}^*}$ states
we obtain from the $X(3915)$,
which we call the $X(3710)$, $X(3820)$ and $X(3955)$ respectively,
see Table \ref{tab:hqs-partners}, should be granted a more conjectural status.
Nevertheless, we notice that the only necessary condition
for the existence of molecular states different than
the $X(3872)$ and $X(4012)$ is that $C_{0b} \geq 0$.

Other theoretical approaches have also predicted several
${\rm D^* \bar{D}^*}$ molecular-like states,
but usually with a mass spectrum incompatible with HQSS.
In the quark model of Ref.~\cite{Maiani:2004vq} there are six
hidden charm diquark-antidiquark states arranged
in a pattern similar to the one we find.
In particular there is a tetraquark $0^{++}$ state at $3723\,{\rm MeV}$
that could be identified with the $X(3710)$ state we obtain.
However, the $2^{++}$ state appears at $3952\,{\rm MeV}$
and is identified with the $X(3940)$ resonance~\cite{Abe:2007jn,Abe:2007sya}.
Unless a considerable violation of HQSS is taking place,
this tensor state is too tightly bound to be considered
the HQSS partner of the $X(3872)$.
Curiously, the two $1^{+-}$ hidden charm diquark-antidiquark states
of Ref.~\cite{Maiani:2004vq} are located at a similar depth below
the $\rm D^* \bar D^*$ and $\rm D \bar D^*$ thresholds respectively
and therefore respect the HQSS expectations.

Another interesting theoretical approach for the study of
hidden charm resonances is the hidden gauge formalism,
using an extension of the $SU(3)$ chiral lagrangians to $SU(4)$ that
implements a particular pattern of $SU(4)$ flavor symmetry breaking.
Within this framework, Gammerman et al.~\cite{Gamermann:2006nm}
have obtained a $0^{++}$ $\rm D\bar D$ molecular state in the vicinity of
$3700\,{\rm MeV}$, that is to be identified with the $X(3710)$
$\rm D \bar D$ molecular state we predict.
The extension of the hidden gauge to axial states~\cite{Gamermann:2007fi}
predicts (among others) a negative C-parity state at $3840\,{\rm MeV}$,
not far way from the $3815-3820\, {\rm MeV}$ mass range we obtain
for the $1^{+-}$ $\rm D\bar D^*$ state.
Finally, the related exploration of resonances generated by the interaction
of two vector mesons in Ref.~\cite{Molina:2009ct}
predicts a series of $0^{++}$, $1^{+-}$ and $2^{++}$
$\rm D^* \bar D^*$ states.
The $0^{++}$ $\rm D^* \bar D^+$ resonance is found
in the region around $3940\,{\rm MeV}$.
Though not identical, this figure does not differ much from the mass of
the $X(3915)$ resonance that we employ as input.
The $1^{+-}$ $\rm D^* \bar D^*$ state of Ref.~\cite{Molina:2009ct}
matches rather well with the mass of the $X(3955)$ state we obtain.
However, the $2^{++}$ ${\rm D^* \bar{D}^*}$ isoscalar resonance 
is considerably different from the $X(4012)$ state:
its mass and width are $M = 3929 \pm 3\,{\rm MeV}$ and
$\Gamma = 29 \pm 10 \,{\rm MeV}$ respectively,
where the dominant decay channel is $\rm D\bar{D}$.
This mass, which is clearly incompatible with the HQSS pattern,
is the result of the remarkably strong vector-vector interaction
generated by the hidden gauge model.
Curiously, the properties of this tensor ${\rm D^* \bar{D}^*}$ resonance
are strikingly similar to those of the $\chi_{c2}(2P)$ charmonium state~\cite{Uehara:2005qd,Aubert:2010ab}
(sometimes referred to as the $Z(3940)$):
$M = 3927.2 \pm 2.6$ and $\Gamma = 24 \pm 6\,{\rm MeV}$,
decaying mostly to $\rm D\bar{D}$~\cite{Nakamura:2010zzi}.

The comparison of the HQSS spectrum with experimentally known states
is however incomplete.
In principle there is so far no experimental evidence in favor (or against)
of the positive C-parity $X(3710)$ and $X(4012)$ states we predict.
Interestingly, the properties of the predicted $X(3955)$ molecular
state are not very different from what is experimentally known about
the $X(3940)$ resonance~\cite{Abe:2007jn,Abe:2007sya},
which have been observed to decay into $\rm D \bar D^*$
(just as would have been expected for a $1^{+-}$ $\rm D^* \bar D^*$ state).
There is a problem though in this identification: the $X(3940)$ is suspected
to be a positive C-parity state, while the $X(3955)$ has negative C-parity.
The reason is that the usual production mechanism
$e^{+}e^{-} \to \gamma^{*} \to J/ \Psi X$ favors the generation
of positive C-parity XYZ states, owing to the quantum numbers
of the intermediate virtual photon and the
final $J / \Psi$.
This mechanism also implies that any prediction about negative C-parity states
will be more difficult to confirm or discard experimentally.
However, even though not so probable, the production of the final $J/\Psi X$
state may happen via two virtual photons, in which case the XYZ
resonance may have negative C-parity.
This alternative production mechanism is not always as suppressed as expected,
as demonstrated in Refs.~\cite{Bodwin:2002fk,Bodwin:2002kk}
for $e^{+}e^{-} \to \gamma^{*}\gamma^{*} \to J/\Psi \, J/\Psi$.
In principle, a similar mechanism could take place in the $X(3940)$ state,
in which case the identification with a $1^{+-}$ $\rm D\bar D^*$ molecule
would be very appealing, but it may also be possible that the $X(3955)$
molecular state has simply not been observed yet.

We have also examined the role played by the OPE potential
in the $\rm D^{(*)} \bar D^{(*)}$ system
(Table~\ref{tab:hqs-partners-OPE-a}).
In agreement with the conclusions of Ref.~\cite{Valderrama:2012},
we have verified that pion exchanges can be treated perturbatively
in the case of isoscalar charm meson-antimeson molecules.
Curiously, the suppression of the OPE effects is larger than naively
expected in terms of the power counting.
The remarkable simplification is that the EFT framework we are proposing
consists entirely on contact range interactions at lowest order.
According to Ref.~\cite{Valderrama:2012}
this interesting simplification will also apply in the isovector
charm sector, as well as the isovector bottom one~\cite{Mehen:2011yh}.
In this regard non-perturbative OPE seems to be only required in the case of
isoscalar bottom meson-antimeson molecules.
This possibility, and the corresponding EFT, was partially explored
in Ref.~\cite{Nieves:2011zz} for the $\rm B\bar{B}^*$/
$\rm B^*\bar{B}$ case.
Lastly, particle coupled channel dynamics are suppressed by two orders
in the counting, as expected, but a more complete analysis would be
welcomed, specially in what regards to regulator dependence.

Even though all the previous states and their possible identification with
theoretically predicted / experimentally known resonances are contingent
on the validity of the molecular hypothesis for the $X(3915)$,
the bottom-line of the approach we advocate is that, provided
we identify at least two molecular states representing
two different combinations of the $C_{0a}$ and $C_{0b}$
counterterms, we will be able to predict the full molecular spectrum
of the $\rm D^{(*)} \bar D^{(*)}$ system.
If the $X(3915)$ identification proves erroneous in the future,
the finding of a different molecular state candidate
could be used to obtain the remaining states.
Owing to the contact range character of the present EFT framework
at lowest order, the calculational effort involved in this task
will be minimal.

\begin{acknowledgments}
We thank E. Ruiz Arriola for discussions.
This research was supported by DGI and FEDER funds, under contract
FIS2011-28853-C02-02, and the Spanish Consolider-Ingenio 2010 Programme
CPAN (CSD2007-00042),  by Generalitat Valenciana under contract
PROMETEO/20090090 and by the EU HadronPhysics2 project,
grant agreement no. 227431. 
\end{acknowledgments}

\appendix
\section{The Effective Lagrangian at Lowest Order}
\label{app:EFT}

In this appendix we write the EFT Lagrangian that describes
the strong interactions of heavy mesons and antimesons
containing a heavy quark $Q$ or antiquark ${\bar Q}$ respectively. 
We use the matrix field $H^{(Q)}$ ($H^{(\bar Q)}$) to denote the following
combination of the pseudoscalar and vector heavy-meson (antimeson) fields
\begin{eqnarray}
H_a^{(Q)} &=& \frac{1+\vsl}2 \left (P_{a\mu}^{* (Q)}\gamma^\mu -
P_a^{(Q)}\gamma_5 \right) \, , \\
H^{(\bar Q)a} &=&  \left (P_{\mu}^{* (\bar Q) a}\gamma^\mu -
P^{(\bar Q)a}\gamma_5 \right) \frac{1-\vsl}2 \, ,
\end{eqnarray}
where the pseudoscalar meson (antimeson) fields are represented by
$P^{(Q)}_a$ ($P^{(\bar Q)}_a$), while $P^{*(Q)}_a$ ($P^{*(\bar Q)}_a$)
is employed for their vector HQSS partners (see, for example,
Ref.~\cite{Falk:1990yz} for further details).
Finally, $v$ is the velocity parameter.
In principle there should be a $v$ subscript to indicate that
we are defining the fields for a specific value of $v$, 
but we have omitted it to avoid complicating the notation. 
The fields are isospin doublets (hence the index a),
where for the pseudoscalar meson and antimesons we have
\begin{eqnarray}
P_a^{(Q / \bar Q)} &=& (P^0,P^+) \, , \\
P_a^{(\bar Q / Q)} &=& (\bar P^0,P^-) \, , 
\end{eqnarray}
plus the analogous expressions for the vector case.
The heavy quark/antiquark superindex changes depending on whether we are
considering charm or bottom meson fields (in the charm case, $D^{0}$ and
$D^{+}$ contain the quark field, while in the bottom case $B^{(0)}$
and $B^{+}$ contain the antiquark field).
The heavy vector meson and antimeson are subjected to the additional condition
\begin{eqnarray}
v \cdot P_{a}^{* (Q)} &=& 0  \, , \\
v \cdot P^{*(\bar Q)a} &=& 0 \, ,
\end{eqnarray}
which in turn defines the three different polarizations
of the heavy vector mesons.

The fields $H_a^{(Q)}$ and $H^{(\bar Q)a}$ respectively transform as a
$(2,\bar 2)$ and $(\bar 2,2)$ representation under
the heavy quark spin $\otimes $ SU(2)$_V$ isospin
symmetry~\cite{Grinstein:1992qt}, that is
\begin{eqnarray}
H_a^{(Q)} &\to& S \left( H^{(Q)} U ^\dagger\right)_a \, , \\
H^{(\bar Q) a} &\to& \left(U  H^{(\bar Q)}\right)^a S^\dagger \, ,
\end{eqnarray}
where $S$ is the heavy quark spin transformation and $U$ the isospin one.
The hermitian conjugate fields are
\begin{eqnarray}
\bar H^{(Q)a} &=& \gamma^0 \, H_a^{(Q)\dagger} \, \gamma^0 \, , 
\end{eqnarray}
\begin{eqnarray}
\bar H_a^{(\bar Q)} &=& \gamma^0 \, \bar H^{(\bar Q)a\dagger} \, \gamma^0 \, ,
\end{eqnarray}
and transform as~\cite{Grinstein:1992qt}
\begin{eqnarray}
\bar H^{(Q)a} &\to&  \left( U \bar H^{(Q)} \right)^a S^\dagger \, ,  \\
\bar H^{(\bar Q)_a} &\to& S\left(\bar H^{(\bar Q)} U^\dagger \right)^a \, .
\end{eqnarray}
Of course, the Lagrangian should be invariant
under the previous symmetry transformations. 

At leading order in the EFT expansion the Lagrangian can be written
as the sum of two contributions
\begin{eqnarray}
{\cal L}^{(0)} = {\cal L}^{(0)}_{4H} + {\cal L}^{(0)}_{\pi HH} 
\end{eqnarray}
where the first one contains a 4-meson interaction vertex and
the second the meson-pion vertex.
The 4-meson contact range Lagrangian consistent with HQSS
and chiral symmetry~\cite{AlFiky:2005jd} reads:
\begin{widetext}
\begin{eqnarray}
{\cal L}_{4H}^{(0)} 
&=& D_{0a} \, {\rm Tr} \left [\bar H^{(Q)a} H^{(Q)}_a \gamma_\mu
  \right] {\rm Tr} \left [ H^{(\bar Q)b}\bar H^{(\bar Q)}_b \gamma^\mu
  \right] % \nonumber \\ && \quad 
+ D_{0b} \, {\rm Tr} \left [\bar H^{(Q)a} H^{(Q)}_a \gamma_\mu \gamma_5
  \right] {\rm Tr} \left [ H^{(\bar Q)b}\bar H^{(\bar Q)}_b \gamma^\mu \gamma_5
  \right] \nonumber \\
&+& E_{0a} \, {\rm Tr}
\left [\bar H^{(Q)a}\,\vec{\tau}^{\,b}_{a}\, H^{(Q)}_b \gamma_\mu \right]
{\rm Tr} \left [ H^{(\bar Q)r}\,\vec{\tau}^{\,s}_{r}\,\bar H^{(\bar
    Q)}_s \gamma^\mu
  \right]  % \nonumber \\ && \quad
+ E_{0b} \, {\rm Tr}
\left [\bar H^{(Q)a} \, \vec{\tau}^{\,b}_{a}\, H^{(Q)}_b 
\gamma_\mu \gamma_5 \right]
{\rm Tr} \left [ H^{(\bar Q)r}\,\vec{\tau}^{\,s}_{r}\,\bar H^{(\bar Q)}_s
\gamma^\mu \gamma_5 \right]  \, . \nonumber \\
\label{eq:L4H}
\end{eqnarray}
\end{widetext}
where $\tau_{ab}$ are the Pauli matrices, and $a$,$b$,$r$ and $s$
are isospin indices.
We notice that for each isospin channel ($I = 0,1$) we have only
two independent constants.

On the other hand, at leading order in the chiral expansion
the HH$\pi$ and $\bar{\rm H} \bar{\rm H} \pi$ couplings
are determined by the Lagrangian~\cite{Grinstein:1992qt}
\begin{eqnarray}
{\cal L}^{(0)}_{\pi HH} &=& -\frac{g}{\sqrt2 f_\pi}
\Big\{ {\rm Tr} \left [\bar H^{(Q)b} H^{(Q)}_a \gamma_\mu \gamma_5\right]
\nonumber \\ && \quad +
{\rm Tr} \left [
  H^{(\bar Q)b}\bar H^{(\bar Q)}_a \gamma^\mu \gamma_5\right] \Big\}
(\vec{\tau} \cdot \partial_\mu  \vec \pi)^{\, a}_b  \nonumber \\ 
&& \quad + \mathcal{O}(\pi^2) \label{eq:LpiHH}
\end{eqnarray}
where $\vec \pi$ is the relativistic field that describes the
pion, $g$ is the $P P^*\pi$ coupling
and $f_{\pi} \simeq 132\,{\rm MeV}$ the pion decay constant.
In the charm sector, $g$ has been determined from the $D^*$ meson decays
in Refs.~\cite{Ahmed:2001xc,Anastassov:2001cw} yielding
$g = 0.59 \pm0.01 \pm 0.07$,
which we approximate by $g \simeq 0.6$.
In the strict heavy quark limit, the latest lattice QCD results suggest
the value $g = 0.449 \pm 0.047 \pm 0.019$,
see Refs.~\cite{Detmold:2011bp,Detmold:2012ge}.
In the normalization above the pion field has dimensions of $[{\rm energy}]$,
while the heavy meson or antimeson fields $H^{(Q)}$ or $H^{(\bar Q)}$ have
dimensions of ${[{\rm energy}]}^{3/2}$:
as usual in heavy quark physics, we employ a non-relativistic normalization
for the heavy mesons that differs from the usual relativistic convention
by a factor of $\sqrt{M_H}$ (see for instance Ref.\cite{Manohar:2000dt}).

\section{Projecting the Potential into the Partial Wave Basis}
\label{app:projecting}

In this appendix we delineate how to project the heavy meson-antimeson
potential into the partial wave basis.
In first place we define the non-relativistic potential for the transition
(not necessarily elastic)
\begin{eqnarray}
{\rm H(1)\bar{H}(2)} \to {\rm H(1')\bar{H}(2')} \, ,
\end{eqnarray}
in terms of the tree level scattering amplitude
\begin{eqnarray}
\mathcal{T}_{\rm tree} = -i\, {\mathcal V}(1+2 \to 1'+2') \, ,
\end{eqnarray}
where $1$,$2$ and $1'$,$2'$ schematically represent
the initial and final state of each of the particles.
For heavy meson-antimeson scattering the initial (final) state is completely
represented by the momentum exchanged between the particles
$\vec{p} = \vec{p}_1 - \vec{p}_2$ ($\vec{p}\,' = \vec{p}_1\,' - \vec{p}_2\,'$)
and by the total and third component of the spin of each of the particles,
which we can collectively call $\sigma = \{ (S_1 m_1) (S_2 m_2) \}$
($\sigma' = \{ (S_1' m_1') (S_2' m_2') \}$).
If we compute ${\mathcal V}(1+2 \to 1'+2')$ in terms of the usual
Feynman rules (each vertex contributes with $i {\cal L}$, additional
$i$ factors for each pion propagator in the case of OPE contributions,
etc..), the relationship between the non-relativistic potential and
the invariant scattering amplitude is
\begin{equation}
\left \langle \vec{p}^{\,\prime}; \sigma' | V | \vec{p}\,;
\sigma \right \rangle = \frac{1}{4} {\mathcal V}
(\vec{p},\sigma \to \vec{p}\,',\sigma') \, .
\label{eq:multipolo1}
\end{equation}
Notice that the factor dividing the invariant scattering amplitude is $4$,
instead of the usual $4 \sqrt{M_1 M_2 M_1' M_2'}$, owing to the $\sqrt{M_H}$
normalization factor included in the heavy meson/antimeson fields.

Now we specify the procedure for the partial wave projection of
the potential.
To denote the different partial waves we employ the spectroscopic notation
$^{2S+1}L_J$, where $S$, $L$ and $J$ are the intrinsic,
orbital and total angular momentum.
With this, we define the states with good angular momentum as follows
\begin{eqnarray}
|p\,; J M L S\rangle &=&  \frac{1}{\sqrt{4\pi}}\,\sum_{M_L, M_S} 
(LSJ|M_L M_S M) \nonumber \\ &\times&
\int d\Omega(\hat p) Y_{L,M_L}(\hat p) |\vec{p}, S M_S\rangle \, ,
\end{eqnarray} 
where $p$ is the modulus of the center-of-mass (c.m.) momentum $\vec{p}$
($= p \times \hat{p}$) of the $\rm H \bar{H}$ pair and
$(L S J | M_L M_S M)$ is a Clebsch-Gordan coefficient.
The normalization of the states above can be determined from the normalization
of the plane wave basis, that is
\begin{eqnarray}
\langle \vec{p}^{\,\prime}; S' M_S'\,|\vec{p}\,; S M_S\rangle 
= (2\pi)^3\,
\delta^3(\vec{p}-\vec{p}^{\, \prime})\,\delta_{S,S'}\delta_{M_S M_S'} \, ,
\nonumber \\
\end{eqnarray}
yielding
\begin{eqnarray}
&& \langle p'; J' M' L' S' \,| p\,; J M L S\rangle = \nonumber \\
&& \phantom{\langle p'; J' M' \,|}
(2\pi)^3\,\frac{\delta(p'-p)}{4 \pi\,p \, p'} \, 
\delta_{JJ'} \delta_{MM'} \delta_{LL'} \delta_{SS'} \, .
\end{eqnarray}
In this basis the partial wave projection of the potential reads
\begin{widetext}
\begin{eqnarray}
V^{S'S}_{JL'L} ( p' ,p) & \equiv &
\langle p'; J M L' S' \,| V |  p\,; J M L S \rangle \nonumber \\
&=& \frac{1}{4\pi}\,
\int d\Omega(\hat p) \int d\Omega(\hat p') \sum_{M_L M_S M_L'M_S'}
(LSJ|M_L M_S M) (L'S'J|M_L' M_S' M) Y^*_{L',M_L'}(\hat p')
Y_{L,M_L}(\hat p) \nonumber \\
&\times& \sum_{m_1 m_2 m_1' m_2'} (S_1 S_2 S| m_1 m_2 M_S)
(S'_1 S'_2 S'| m_1' m_2' M_S') \left \langle \vec{p}^{\,\prime};
(S'_1 m'_1) (S'_2 m'_2) | V | \vec{p}\,;
(S_1 m_1) (S_2 m_2) \right \rangle \, ,
\label{eq:multipolo2}
\end{eqnarray}
\end{widetext}
where thanks to rotational invariance the above matrix element is
independent of the third component of
the total angular momentum $M$.

\section{The Lowest Order Heavy Meson-Antimeson Potential
and Its Partial Wave Projection} 
\label{app:LO-pot}

The lowest order $\rm H \bar H$ potential contains a contact
and a finite range (pion exchange) piece.
The EFT potential can be derived from the tree level scattering
amplitudes resulting from the ${\cal L}^{(0)}_{4H}$ and
${\cal L}^{(0)}_{\pi HH}$ Lagrangians
of Eqs.~(\ref{eq:L4H}) and (\ref{eq:LpiHH}).
Even though the partial wave projection of the contact piece is trivial,
we will start with the OPE potential in order to fix the notation.
The tree level invariant amplitude can be obtained from the pion-meson
${\cal L}^{(0)}_{\pi HH}$ Lagrangian of Eq.~(\ref{eq:LpiHH}),
taking the schematic form in the strict heavy quark limit,
\begin{eqnarray}
\label{eq:OPE-amplitude}
{\mathcal V}_{\rm OPE}(\vec{p}^{\,\prime}, \vec{p}\,) \propto
\frac{(\vec{a}\cdot\vec{q}\,)(\vec{b}\cdot\vec{q}\,)}{\vec{q}^{~2}+m_\pi^2}
\, ,
\end{eqnarray}
where $\vec{q}=\vec{p}-\vec{p}^{\, \prime}$ (that is, the potential is
local) and $\vec{a}$, $\vec{b}$ is the corresponding polarization
operator in each of the ${\rm H \bar H} \pi$ vertices. The
proportionality factor is $g^2 / 8 f_{\pi}^2$ times a sign that
depends the pseudoscalar or vector nature of each of the particles in
the initial and final states.  The modifications to take into account
in the above equation the mass difference between the pseudoscalar and
vector heavy mesons masses are also discussed in
Refs.~\cite{Nieves:2011zz,Valderrama:2012}.  Since we will not be
considering particle coupled channels with the OPE interaction, this
becomes an issue in this work only for the $D\bar D^*\to D^*\bar D$
channel. In that case in Eq.~(\ref{eq:OPE-amplitude}), $m_\pi^2$
should be substituted by $\mu_\pi^2= m_\pi^2-(m_{D^*}-m_D)^2$. Indeed,
$\mu_\pi^2 \le 0$, since there is a very small absorptive contribution
from the $D\bar D\pi$ channel. We will neglect it, as in
Ref.~\cite{Nieves:2011zz}, and in that case we will consider only the
real part of the potential.
We now continue by Fourier transforming the amplitude above
into coordinate space.
This step, though counter-intuitive at first sight, will enormously
facilitate the calculation of the partial wave projection of
the potential in momentum space.
We remind that the coordinate and momentum space potentials are related by
\begin{equation}
\langle \vec{p}^{\,\prime}; \sigma' | V | \vec{p}\,; \sigma \rangle=
\int {d^3 r}\,
e^{i(\vec{p}-\vec{p}^{\,\prime})\cdot\vec{r}}
\left \langle \sigma' | V(\vec{r}\,) | 
\sigma \right \rangle \, ,
\end{equation}
and we use the symbols $\sigma$ and $\sigma'$ to encode
all the spin indices (see the previous appendix).
Then we make use of a well-known relationship
\begin{eqnarray}
&& \int \frac{d^3 q}{(2 \pi)^3} \,
\frac{(\vec{a}\cdot\vec{q}\,)(\vec{b}\cdot\vec{q}\,)}{\vec{q}^{~2}+m_\pi^2}
e^{-i\vec{q}\cdot \vec{r}} \nonumber \\
&& \phantom{HolaJuan} = - \vec{a} \cdot \vec{\nabla} \, 
\vec{b} \cdot \vec{\nabla} \, 
\left( \frac{e^{-m_\pi r}}{4\pi\,r}\right) 
\end{eqnarray}
from which we obtain the form of the OPE potential 
\begin{eqnarray}
V_{OPE}(\vec{r}) \propto \frac{\vec{a}\cdot\vec{b}}{3}\,\delta^3(\vec{r}) 
- \Big( v_C(r)\,\vec{a}\cdot\vec{b} 
+ \, v_T(r)\,S_{12}(\vec{a},\vec{b}) \Big) \, . \nonumber \\
\end{eqnarray}
In this equation $S_{12}(\vec{a},\vec{b})$ is the tensor operator,
which we define as 
\begin{eqnarray}
S_{12}(\vec{a},\vec{b}) &=&
\frac{3(\vec{a}\cdot\vec{r}\,)(\vec{b}\cdot\vec{r}\,)}{r^2}
- \vec{a}\cdot\vec{b} \, .
\end{eqnarray}
In turn, the central and tensor pieces of the potential,
$v_c$ and $v_T$, are given by
\begin{eqnarray} 
v_C(r) &=& \frac{m_\pi^3}{12 \pi }
\left( \frac{e^{-m_\pi r}}{m_\pi r}\right) \, , \\
v_T(r) &=& v_C(r)\,
\left (1 +\frac{3}{m_\pi r} + \frac{3}{(m_\pi r)^2}  \right) \, .
\end{eqnarray}
Of course, we are interested in the partial wave projection of the potential
above, $V^{S'S}_{JL'L}(r)$, since its partial wave Fourier transform
provides the multipole expansion of the potential in momentum space
(Eq.~(\ref{eq:multipolo2}))
\begin{eqnarray}
V^{S'S}_{JL'L} ( p',p) &=& 4\pi i^{L-L'} \nonumber \\ &\times& 
\int_0^{+\infty} dr \, r^2 j_L(p\,r) \, j_{L'}(p'r) \, V^{S'S}_{JL'L}(r)
\, , \nonumber \\ \label{eq:cross-projection}
\end{eqnarray}
where $j_L(x)$ represent the spherical Bessel function of order $L$.
The advantage of this expression is that it can be analytically evaluated
with relative ease for the OPE potential.
After a bit of Racah algebra and taking into account all the signs and factors
we have obviated so far (again, details can be consulted
in Ref.~\cite{Valderrama:2012}), we arrive at the final
expression for the OPE potential in the partial wave basis
\begin{eqnarray}
{(V_{OPE})}_{J\,L'L}^{S'S}(r) &=& - \frac{g^2}{6f_\pi^2}\,
\vec{\tau}_1\cdot \vec{\tau}_2
\,\frac{\delta(r)}{4 \pi r^2} \,{C}_{12}
\nonumber \\ && + \frac{g^2}{2f_\pi^2}\,\vec{\tau}_1\cdot \vec{\tau}_2
\left[ v_C(r) \, {C}_{12} +  v_T(r) \, {S}_{12} \right] \, , \nonumber \\
\label{eq:OPE-projected}
\end{eqnarray}
where all the calculational complications are conveniently hidden
in $C_{12}$ and $S_{12}$, the partial wave projections of
the $\vec{a} \cdot \vec{b}$ and $S_{12}(\vec{a},\vec{b})$ operators:
$C_{12}$ and $S_{12}$ depend on $J$, $L$, $L'$, $S$ and thus encode all
the information required to determine the coordinate space potential
in a particular partial wave.

The most compact way to write the $C_{12}$ and $S_{12}$ factors is
in matrix form, where the matrix is defined in a basis formed
by the set of partial waves with well-defined quantum numbers $J^{PC}$.
We can illustrate this by considering all the $J^{PC}$ combinations
that contain S-waves, that is, the $J^{PC}$ values we have studied
in this work.
First we consider the set of $0^{++}$ partial waves, defined as
\begin{eqnarray}
{\mathcal B}(0^{++}) =
\Big \{ D\bar D(^1S_0), D^*\bar D^*(^1S_0), D^* 
\bar D^*(^5D_0)  \Big \} \, , \nonumber \\
\end{eqnarray}
from which the ${\bf C}_{12}$ and ${\bf S}_{12}$ matrices are
\begin{eqnarray}
{\bf C}_{12}(0^{++}) &=& \left( \begin{matrix}
0 & &  \cr 
-\sqrt{3}& 2&  \cr
0 &0 &-1
\end{matrix}
\right) \, , \\
{\bf S}_{12}(0^{++}) &=& 
\left( \begin{matrix}
0 &&  \cr 
0& 0&  \cr
\sqrt{6} &\sqrt{2} &2
\end{matrix}\right) \, , 
\end{eqnarray}
where we have only shown the lower and diagonal components
as the matrices are symmetric.
Next we move to the $1^{++}$ case, for which we have the particle states
$\rm D \bar D^*$ and $\rm D^* \bar D$ that we need to arrange in states
with good C-parity.
We thus define 
\begin{eqnarray}
[D\bar D^*(\eta)] = 
\frac{1}{\sqrt{2}} \, \left[ D\bar D^* - \eta \, D^*\bar D \right ] \, .
\end{eqnarray}
In this convention the intrinsic $C-$parity of these states
is independent of the isospin and equal to $\eta$.
The $1^{++}$  basis thus reads
\begin{eqnarray}
{\mathcal B}(1^{++}) &=& \Big \{ 
[D\bar D^*(+)](^3S_1), [D\bar D^*(+)](^3D_1), \nonumber \\
&& \phantom{p} D^* \bar D^*(^5D_1)  \Big \} \, , 
\end{eqnarray}
for which we obtain the matrices
\begin{eqnarray}
{\bf C}_{12}(1^{++}) &=& \left( \begin{matrix}
-1 & &  \cr 
0& -1&  \cr
0 &0 &-1
\end{matrix}
\right) \, , \\
{\bf S}_{12}(1^{++}) &=& \left( \begin{matrix}
0 &&  \cr 
\sqrt{2}& -1&  \cr
\sqrt{6} &\sqrt{3} &1
\end{matrix} \right) \, , 
\end{eqnarray}
The next case is $1^{+-}$, in which we employ the basis
\begin{eqnarray}
\mathcal{B}(1^{+-}) &=& \Big \{ [D\bar D^*(-)](^3S_1), [D\bar D^*(-)](^3D_1), 
\nonumber \\ && \phantom{p}
D^* \bar D^*(^3S_1) , D^* \bar D^*(^3D_1)  \Big \} \, ,
\end{eqnarray}
and get the matrices
\begin{eqnarray}
{\bf C}_{12}(1^{+-}) &=&
\left( \begin{matrix}
1 & & & \cr 
0& 1& &  \cr
-2 &0 &1 & \cr
0 & -2 & 0& 1
\end{matrix}
\right) \,  , 
\end{eqnarray}
\begin{eqnarray}
{\bf S}_{12}(1^{+-}) &=& 
\left( \begin{matrix}
0 &&&  \cr 
-\sqrt{2}& 1& &  \cr
0 &-\sqrt{2} &0\cr
-\sqrt{2}& 1&-\sqrt{2}&1
\end{matrix}\right)  \, .
\end{eqnarray}

\begin{widetext}
The most complex case is $2^{++}$, in which OPE mixes all
the possible particle channels.
We thus work in the basis
\begin{eqnarray}
{\mathcal B}(2^{++}) = \Big \{ D\bar D (^1D_2), [D\bar D^*(+)](^3D_2), 
D^* \bar D^*(^1D_2) , D^* \bar D^*(^5S_2) , D^* \bar D^*(^5D_2),
D^* \bar D^*(^5G_2) \Big \} \, ,
\end{eqnarray}
and obtain the matrices
\begin{eqnarray}
{\bf C}_{12}(2^{++}) &=& \left( \begin{matrix}
0        &  &  &  &  &  &\cr 
0        &-1&  &  &  &  &\cr
-\sqrt{3}&0 &2 &  &  &  &\cr
0        &0 &0 &-1&  &  &\cr
0        &0 &0 &0 &-1&  &\cr
0        &0 &0 &0 & 0&-1&
\end{matrix}
\right) \, , \\
{\bf S}_{12}(2^{++}) &=& \left( \begin{matrix}
0                 &                    &                  &  &  &  &\cr 
0                 &1                   &                  &  &  &  &\cr
0                 &0                   &0                 &  &  &  &\cr
\sqrt{\frac65}    &-3\sqrt{\frac25}    &\sqrt{\frac25}    &0 &  &  &\cr
-2\sqrt{\frac37}  &\frac3{\sqrt{7}}    &-\frac2{\sqrt{7}} &-\sqrt{\frac{14}5}&-\frac37&  &\cr
6\sqrt{\frac3{35}}&\frac{12}{\sqrt{35}} &\frac6{\sqrt{35}} &0&-\frac{12}{7\sqrt{5}} &\frac{10}7&
\end{matrix}
\right) \, .
\end{eqnarray}
\end{widetext}

Now we consider the  contact range piece of the potential, which can be
easily derived from the 4-meson ${\cal L}^{(0)}_{4H}$ Lagrangian
of Eq.~(\ref{eq:L4H}).
For that, we expand the ${\cal L}^{(0)}_{4H}$ Lagrangian into its explicit
representation in terms of pseudoscalar
and vector heavy meson fields
(detailed expressions can be consulted in Ref.~\cite{Valderrama:2012}).
The subsequent partial wave projection is straightforward 
(owing to the simplification of working with S-waves only):
\begin{eqnarray}
{(V_C)}_{J\,L'L}^{S'S}(r) &=&  \frac{\delta(r)}{4 \pi r^2}\,(D_{0a} - 
E_{0a}\,\vec{\tau}_1\cdot \vec{\tau}_2)\,\delta_{J S}\,\delta_{J' S'}\,
\delta_{L 0}\,\delta_{L' 0} \nonumber \\
&+&  \frac{\delta(r)}{4 \pi r^2}\,(D_{0b} - 
E_{0b}\,\vec{\tau}_1\cdot \vec{\tau}_2)\,C_{12} \, ,
\label{eq:contact-projected}
\end{eqnarray}
from which the potentials of Eqs.~(\ref{eq:contact1})--(\ref{eq:contact2-b})
are derived,
Notice however that in Eqs.~(\ref{eq:contact1})--(\ref{eq:contact2-b})
we have specified the potential according to the $J^{PC}$ quantum number
of the heavy meson-antimeson system.
The relation between the couplings of
Eqs.~(\ref{eq:contact1})--(\ref{eq:contact2-b}) and the corresponding ones
in the ${\cal L}^{(0)}_{4H}$ Lagrangian is provided by the expressions
\begin{eqnarray}
C_{0a} &=& D_{0a} + 3\,E_{0a} \, , \nonumber \\ 
C_{0b} &=& D_{0b} + 3\,E_{0b} \, 
\end{eqnarray}
where we have isolated the isoscalar contribution~\footnote{For the isovector
states we get $C_{0a}(I=1) = D_{0a} - E_{0a}$ and 
$C_{0b}(I=1) = D_{0b} - E_{0b}$}. 
If we are considering the full potential resulting from the sum of the contact
range and OPE contribution
\begin{eqnarray}
{V}_{J\,L'L}^{S'S}(r) = {(V_C)}_{J\,L'L}^{S'S}(r) + {(V_{OPE})}_{J\,L'L}^{S'S}(r) \, ,
\end{eqnarray}
we see that the $\delta(r)$ contribution within the OPE potential
can be absorbed within the contact range piece by means of
the replacement
\begin{equation}
C_{0b} \to C_{0b}-\frac{g^2}{2f_\pi^2} \label{eq:c0bchange} \, ,
\end{equation}
which in fact is able to account for the bulk of the change of this operator
when we add the OPE potential, see the related discussion
in Sect.~\ref{sec:subleading}.

%\bibliography{hqss-x3872}
%Merlin.mbs v4.21 2009-07-09.

%

\end{document}